\documentclass[manuscript]{aastex}

\def\Ga {G28.87$+$0.07}
\def\Gb {G16.59$-$0.05}
\def\Gc {G23.01$-$0.41}
\def\wat {H$_{2}$O}
\def\meth {CH$_{3}$OH}
\def\kms {km s$^{-1}$}
\def\mum {$\mu$m}

\shorttitle{VLBI study of maser kinematics} \shortauthors{Li et al.}

\begin{document}

\title{Massive star-formation toward
\Ga\ (IRAS\,18411$-$0338) investigated by means of maser kinematics
and radio to infrared, continuum observations}

\author{J. J. Li\altaffilmark{1,2}, L. Moscadelli\altaffilmark{2},
R. Cesaroni\altaffilmark{2}, R. S. Furuya\altaffilmark{3}, Y.
Xu\altaffilmark{1}, T. Usuda\altaffilmark{3}, K. M.
Menten\altaffilmark{4}, M. Pestalozzi\altaffilmark{5}, D.
Elia\altaffilmark{5}, and E. Schisano\altaffilmark{5}}

\altaffiltext{1}{Purple Mountain Observatory, Chinese Academy of
Sciences, Nanjing 210008, China; jjli@pmo.ac.cn}
\altaffiltext{2}{INAF-Osservatorio Astrofisico di Arcetri, Largo E.
Fermi 5, 50125 Firenze, Italy} \altaffiltext{3}{Subaru Telescope,
National Astronomical Observatory of Japan, 650 North A'{o}hoku
Place, Hilo, HI 96720, USA} \altaffiltext{4}{Max-Planck-Institut
f\"{u}r Radioastronomie, Auf dem H\"{u}gel 69, 53121 Bonn, Germany}
\altaffiltext{5}{INAF-Istituto Fisica Spazio Interplanetario, Via
Fosso del Cavaliere 100, I-00133 Roma, Italy}

\begin{abstract}

We used the Very Long Baseline Array (VLBA) and the European VLBI
Network (EVN) to perform phase-referenced VLBI observations of the
three most powerful maser transitions associated with the high-mass
star-forming region \Ga: the 22.2~GHz \wat, 6.7~GHz \meth, and
1.665~GHz OH lines. We also performed VLA observations of the radio
continuum emission at 1.3~and~3.6~cm and Subaru observations of the
continuum emission at 24.5~\mum. Two centimeter continuum sources
are detected and one of them (named ``HMC'') is compact and placed
at the center of the observed distribution of \wat, \meth\ and OH
masers. The bipolar distribution of line-of-sight (l.o.s) velocities
and the pattern of the proper motions suggest that the water masers
are driven by a (proto)stellar jet interacting with the dense
circumstellar gas. The same jet could both excite the centimeter
continuum source named ``HMC'' (interpreted as free-free emission
from shocked gas) and power the molecular outflow observed at larger
scales -- although one cannot exclude that the free-free continuum
is rather originating from a hypercompact \ion{H}{2} region. At
24.5~\mum, we identify two objects separated along the north-south
direction, whose absolute positions agree with those of the two VLA
continuum sources. We establish that $\sim$90\% of the luminosity of
the region ($\sim$2$\times$10$^{5}~L_\sun$) is coming from the radio
source ``HMC'', which confirms the existence of an embedded massive
young stellar object (MYSO) exciting the masers and possibly still
undergoing heavy accretion from the surrounding envelope.
\end{abstract}


\section{INTRODUCTION}

Understanding the process of high-mass star formation represents a
challenge from both a theoretical and observational point of view.
By now, three main competing concepts of massive star formation have
been discussed in the recent literature \citep[see][and references
therein]{zin07}: formation through coalescence of lower-mass stars
in extremely dense stellar clusters \citep{bon98}; monolithic
collapse of isolated, turbulence supported cores \citep{mck03}; and
competitive accretion where the most massive protostars benefit of
the gravitational well created by the lower-mass stars surrounding
them, thus boosting the accretion rate \citep[e.g.,][]{bon06}. The
available observational evidence is still inadequate to discern
between these different scenarios, mostly because observations of
massive young stellar objects (MYSOs) are complicated by the large
distances of typically several kpc, implying small angular scales,
and the fact that such objects are born still deeply enshrouded in
dense dusty envelopes, opaque to optical and near-infrared
wavelength radiation.

A possible way to distinguish between different models is to explore
the physical conditions and the dynamical processes of the gas close
to the protostars. However, present (sub)millimeter interferometers
and infrared (IR) telescopes attain angular resolutions
$\geq$0\farcs25, corresponding to thousands of astronomical units
(AU) at the typical distances of MYSOs of several kpc. This is
insufficient to investigate the kinematics of the gas in the
immediate surroundings of the forming star, where crucial phenomena
such as accretion, rotation, and ejection are likely most prominent.
The above mentioned limitation can be bypassed by Very Long Baseline
Interferometry (VLBI) observations of maser lines, which attain
linear resolutions of a few AU.

A few years ago, we started an observational project to study the
high-mass star-forming process by comparing interferometric images
of thermal lines of molecular tracers with multi-epoch VLBI studies
in three well known maser species (OH, \meth\ and \wat). At present,
three sources (IRAS\,20126$+$4104, \Gb\ and \Gc) have been already
analyzed \citep{mos11,san10a,san10b}. The results demonstrate that
the synergy between VLBI, multi-species maser observations and
(sub)mm interferometric observations of thermal molecular lines is
crucial to achieve a multi-scale picture of the environment of newly
formed massive (proto)stars. In the present article we report on our
VLBI maser study of the high-mass star-forming region (HMSFR) \Ga,
and compare our results with available complementary interferometric
data of continuum and thermally excited molecular line emission, as
well as with sub-arcsec resolution mid-IR images obtained with the
Cooled Mid-Infrared Camera and Spectrometer (COMICS) on the Subaru
telescope.

In Section~2, we provide a review of previous observations toward
this region. Section~3 describes the VLBI observations of the maser
transitions, the Very Large Array (VLA) observations of the radio
continuum emission at 1.3~cm and 3.6~cm, and the mid-IR observations
performed with the Subaru/COMICS telescope. In Section~4, we present
the distributions of the three types of masers, the water maser
proper motions, and the properties of the radio continuum. The
results are discussed in Section~5 and the conclusions are drawn in
Section~6.

\section{THE HMSFR \Ga}

The HMSFR \Ga\ (alias IRAS\,18411$-$0338) is located at a kinematic
distance of 7.4~kpc\footnote{Since the ambiguity between the near
($\approx$6~kpc) and the far ($\approx$9~kpc) kinematical distance
is not resolved \citep{sew04}, we adopt the distance to the tangent
point.} \citep{cod97, fur08} and is associated with a pc-scale clump
with a mass in excess of 10$^3~M_\sun$ and a bolometric luminosity
of $\sim$2$\times$10$^5~L_\sun$ \citep{lop09, fau04}. From the
observation of NH$_3$, the systemic velocity of the region ($V_{\rm
sys}$) with respect to the local standard of rest (LSR) is found to
be 101.0~\kms\ \citep{cod97}.

On a large scale, mid-IR observations at 8, 10.5, and 18.1~\mum\
have resolved the IRAS source in two objects along the north-south
direction \citep[Figure~1]{bui05, fur08}. 
Both sources could be associated with an Midcourse Space Experiment
(MSX) point source \citep{ega03} and a 2.2~\mum\ source with strong
near-infrared excess, lying 3\arcsec--4\arcsec\ to the southeast of
the northern, brighter mid-IR object \citep{tes94}. Bolometer
observations at 1.2~mm \citep{fau04} have detected a peak of dust
continuum emission associated with the two mid-IR sources and the
near-IR source. Single-dish maps in the CS(5--4) and C$^{18}$O(2--1)
lines have revealed an arcmin-scale molecular clump and a monopolar
(red-shifted) outflow is seen in the $^{13}$CO(2--1) line
\citep{shi03,lop09}. Single-dish surveys have detected emission in
dense gas tracers as CS(7--6), CS(2--1), and CS(1--0)
\citep{plu92,bro96,ang96}, and in the H110$\alpha$ radio
recombination line \citep{sew04,ara07}, which suggests the presence
of a diffuse \ion{H}{2} region. In addition, OH maser emission has
been found in the main (1.665 and 1.667~GHz) hyperfine structure
lines and in the 1.720~GHz satellite line, whereas the 1.612~GHz OH
satellite has been observed in absorption \citep{lit82,cas83}. Weak
\meth\ and strong \wat\ maser emission have been reported at 6.7~GHz
\citep{cas95} and 22.2~GHz \citep{gen79}, respectively.

Interferometric observations at millimeter wavelengths have revealed
multiple sources in this region \citep[Figure~1]{fur08} and show
that the dust continuum peak coincides with the position of the
northern, stronger mid-IR source. The mass (100~$M_\sun$) from the
3~mm dust continuum emission and the rotational temperature (93~K)
inferred from the CH$_{3}$CN lines suggest that one is dealing with
a hot molecular core (HMC). In this study, we will focus on this
HMC, as it contains all the tracers of high-mass star formation that
we have investigated with our observations, namely the masers and
the free-free and mid-IR continuum sources.  As shown by
\cite{cod97} and \cite{fur08}, the HMC is seen in NH$_{3}$,
HCO$^{+}$, and CH$_{3}$CN line emission. The observations of
\cite{cod97} were not sensitive enough to detect continuum emission
at 1.3~cm, whereas our new images demonstrate the existence of a
free-free continuum source inside the HMC (see below).  A
wide-angle, massive ($\geq$100~$M_\sun$) bipolar outflow has been
detected in the $^{12}$CO(1--0) line, oriented perpendicular to a
CH$_{3}$CN velocity gradient detected in the HMC \citep{fur08}. The
absolute positions of the 1.665~GHz OH and 22.2~GHz \wat\ masers,
derived by \citet{for99} using the VLA with an accuracy of 0\farcs5,
indicate that these maser emissions are associated with the HMC.

All these features are consistent with the presence of a deeply
embedded MYSO(s) inside the HMC, possibly accompanied by lower mass
objects lying in the neighborhoods of the HMC itself.

\section{OBSERVATION AND DATA REDUCTION}

\subsection{VLA: 1.3~cm and 3.6~cm Continuum}

We observed the HMSFR \Ga\ (tracking center: R.A.(J2000) = $18^{\rm
h}43^{\rm m}46.24^{\rm s}$ and Dec.(J2000) = $-03\degr35\arcmin
30.4\arcsec$) with the VLA at X (3.6~cm) and K (1.3~cm) bands in
both the C-array (in 2008 April) and A-array configurations (in 2008
November). At 3.6~cm in both configurations and at 1.3~cm in the
C-array configuration, the continuum mode of the correlator was
used, resulting in an effective bandwidth of 172~MHz. At 1.3~cm in
the A-array we used mode ``4'' of the correlator, with a pair of
intermediate frequency (IF) bands, each with a width of 3.125~MHz
(64~channels) centered on the strongest \wat\ maser line and a pair
of 25~MHz wide IF bands (8~channels each) sufficiently offset from
the maser lines to obtain a measurement of the continuum emission.
The two bands centered at the same frequency (detecting both
circular polarizations) were averaged to decrease the noise.

At X-band, 3C\,286 (5.2~Jy) and 3C\,48 (3.1~Jy) were used as primary
flux calibrators, while 1832$-$105 (1.4~Jy) was the phase
calibrator. For the K-band observations, the primary flux calibrator
was 3C\,286 (2.5~Jy), the phase calibrator 1832$-$105 (1.0~Jy), and
the bandpass calibrator (for the A-array only) 1733$-$130 (3.7~Jy).

The data were calibrated with the NRAO\footnote{The National Radio
Astronomy Observatory is operated by Associated Universities, Inc.,
under cooperative agreement with the National Science Foundation.}
Astronomical Image Processing System (AIPS) software package using
standard procedures. Only for the A-array data at 1.3~cm, several
cycles of self-calibration were applied to the strongest maser
channel, and the resulting phase and amplitude corrections were
eventually transferred to all the other line channels and to the
K-band continuum data \citep{men95}. This procedure resulted in a
significant (at least a factor 2) improvement in the signal-to-noise
ratio (SNR).

Natural-weighted maps were made with task IMAGR of AIPS for both the
continuum and the line data. Simultaneous observation of the line
and continuum emission at K-band made it possible to obtain the
relative position of the continuum image with respect to the \wat\
maser spots with great precision ($\sim$0\farcs1). Finally, we
identified the maser spots that appear in both our Very Long
Baseline Array (VLBA) and VLA images and, using the accurate
absolute position information obtained from the VLBA observations,
we re-centered the VLA spots and continuum image. The correction
thus applied turns out to be $<$20~mas in each coordinate. We
conclude that the absolute astrometrical precision of the continuum
maps presented in this paper is $\sim$0\farcs1.

\subsection{VLBI: Maser Lines}
\label{mas_obs}

We conducted VLBI observations of the \wat\ and \meth\ masers (at
four epochs) and of the OH maser (at a single epoch) toward \Ga\ at
22.2~GHz, 6.7~GHz, and 1.665~GHz, respectively. We give more
technical details on the observations in the following subsections.
In order to determine the masers' absolute positions, we performed
phase-referencing observations in fast switching mode between the
maser source and the calibrator J1834$-$0301. This calibrator has an
angular offset from the maser source of 2\fdg4 and belongs to the
list of sources of the VLBA Calibrator Survey 3 (VCS3). Its absolute
position is known with a precision better than $\pm$1.5~mas, and its
flux measured with the VLBA at S and X bands is 154 and
156~mJy~beam$^{-1}$, respectively. Five fringe finders
(J1642$+$3948; J1751$+$0939; J1800$+$3848; J2101$+$0341;
J2253$+$1608) were observed for bandpass, single-band delay, and
instrumental phase-offset calibration.

The data were reduced with AIPS following the VLBI spectral line
procedures. For a description of the general data calibration and
the criteria used to identify individual masing clouds, derive their
parameters (position, intensity, flux and size), and measure their
(relative and absolute) proper motions, we refer to the recent paper
on VLBI observations of \wat\ and \meth\ masers by \citet{san10a}.
For each VLBI epoch, \wat\ maser absolute positions have been
derived by employing two methods, direct phase-referencing and
inverse phase-referencing, and the two procedures always gave
consistent results. \meth\ and OH maser absolute positions have been
derived only by using direct phase-referencing. The derived absolute
proper motions have been corrected for the apparent proper motion
due to the Earth's orbit around the Sun (parallax), the solar motion
and the differential Galactic rotation between our LSR and that of
the maser source. We have adopted a flat Galactic rotation curve
($R_0 = 8.3\pm0.23$~kpc, $\Theta_0 = 239\pm7 $~\kms) \citep{bru11},
and the solar motion of $U = 11.1 ^{+0.69}_{-0.75}$, $V = 12.24
^{+0.47}_{-0.47}$, and $W = 7.25 ^{+0.37}_{-0.36}$~\kms\ by
\citet{sch10}, who recently revised the Hipparcos satellite results.

\subsubsection{VLBA: 22.2~GHz \wat\ Masers}

We observed the HMSFR \Ga\ (tracking center: R.A.(J2000) = $18^{\rm
h}43^{\rm m}46.22^{\rm s}$ and Dec.(J2000) = $-03\degr35\arcmin
29.9\arcsec$) with the VLBA in the $6_{16}-5_{23}$ \wat\ transition
(rest frequency 22.235079~GHz). The observations (program code:
BM244) consisted of 4 epochs: 2006 April 23, 2006 June 30, 2006
September 28, and 2007 January 18. During a run of about 6~h per
epoch, we recorded the dual circular polarization with 16~MHz
bandwidth centered at an LSR velocity ($V_{\rm LSR}$) of 101.0 \kms.
The data were processed with the VLBA FX correlator in Socorro (New
Mexico) using an averaging time of 1~s and 1024 spectral channels.

Images were produced with natural weighting, cleaned and restored
with an elliptical Gaussian with full width at half maximum (FWHM)
size of about 1.3~mas $\times$ 0.6~mas at a position angle (PA) of
$-$4\degr\ (east of north), with small variation from epoch to epoch
(shown in Table~\ref{tbl:mas-abs-pos}). The interferometer
instantaneous field of view was limited to $\sim$2\farcs7. In each
observing epoch the on-source integration time was $\sim$2.5~h
resulting in an effective rms noise level of the channel maps
($\sigma$) in the range 0.006--0.015~Jy~beam$^{-1}$. The spectral
resolution was 0.2~\kms.

\subsubsection{EVN: 6.7~GHz \meth\ Masers}

We observed \Ga\ (tracking center: R.A.(J2000) = $18^{\rm h}43^{\rm
m}46.22^{\rm s}$ and Dec.(J2000) = $-03\degr35\arcmin 29.9\arcsec$)
with the European VLBI Network (EVN) in the $5_{1}-6_{0}A^{+}$
\meth\ transition (rest frequency 6.668519~GHz). Data were taken in
4 epochs (program code: EM061 and EM069) separated by $\sim$1~year:
2006 February 28, 2007 March 18, 2008 March 18, and 2009 March 17.
In the first two epochs, the antennae involved in the observations
were Cambridge, Jodrell2, Effelsberg, Hartebeesthoek, Medicina,
Noto, Torun, and Westerbork. Since we found out that the longest
baselines involving the Hartebeesthoek antenna (e.g., the Ef-Hh
baseline is $\sim$8042~km) heavily resolve the maser emission and do
not produce fringe-fit solutions, the Hartebeesthoek antenna was
replaced with the Onsala antenna in the third epoch. During a run of
$\sim$6~h, we recorded the dual circular polarization in two IF
bands of width of 2~MHz and 16~MHz, both centered at an LSR velocity
of 101.0~\kms. The 16~MHz band was used to increase the SNR of the
weak continuum calibrator. The data were processed with the MKIV
correlator at the Joint Institute for VLBI in Europe
(JIVE-Dwingeloo, the Netherlands) using an averaging time of 1~s and
1024 spectral channels for each observing bandwidth.

Images were produced with natural weighting, cleaned and restored
with an elliptical Gaussian with FWHM size of about 12~mas $\times$
6~mas at a PA of 21\degr, with small variations from epoch to epoch
(shown in Table~\ref{tbl:mas-abs-pos}). The interferometer
instantaneous field of view was limited to $\sim$9\farcs2.
Using an on-source integration time of $\sim$2.2~h, the
effective rms noise level of the channel map ($\sigma$) varied in
the range 0.005--0.01~Jy~beam$^{-1}$. The 2~MHz band spectral
resolution was 0.09~\kms.

\subsubsection{VLBA: 1.665~GHz OH Masers}

We observed \Ga\ (tracking center: R.A.(J2000) = $18^{\rm h}43^{\rm
m}46.34^{\rm s}$ and Dec.(J2000) = $-03\degr35\arcmin 29.9\arcsec$)
with the VLBA in the $^{2}\Pi_{3/2} J = 3/2$ OH transition (rest
frequency 1.665401~GHz) on 2007 April 28 (program code: BM244O).
During a run of $\sim$6~h, we recorded the dual circular
polarization with two IF bands of width of 1~MHz and 4~MHz, both
centered at $V_{\rm LSR}=103.0$~\kms. The 4~MHz bandwidth was used
to increase the SNR of the weak L-band signal of the continuum
calibrator. The data were processed with the VLBA FX correlator in
two correlation passes using 1024 and 128 spectral channels for the
1~MHz and 4~MHz bands, respectively. In each correlator pass, the
data averaging time was 2~s.

Images were produced with natural weighting, cleaned and restored
with an elliptical Gaussian with FWHM size of 17.8~mas $\times$
10.9~mas at a PA of 1\degr. The interferometer instantaneous field
of view was limited to $\sim$18\farcs5. The
on-source integration time was $\sim$1.9~h, resulting in an
effective rms noise level in each velocity channel of
$\sim$0.02~Jy~beam$^{-1}$. The spectral resolution was 0.2~\kms.

\subsection{Subaru/COMICS: 24.5~\mum\ Mid-infrared Continuum}

Using the mid-infrared imaging spectrometer \citep[COMICS;][]{kat00}
at the Cassegrain focus of the 8.2~m Subaru Telescope, we carried
out imaging observations of the 24.5~\mum\ emission toward \Ga\ on
2008 July 15. For this purpose, we configured the Q24.5 filter, and
employed chopping mode for subtracting the sky-background emission.
The camera provides a field of view of $\sim 42\arcsec\times
32\arcsec$ with a pixel size of 0\farcs13. Flux calibration was
performed towards four standard sources listed in \citet{coh99}:
HD146051, HD186791, HD198542, and HD198542. We estimated the overall
uncertainty in the flux calibrations to be less than 10\%.
Astrometric calibration was performed by smoothing the 24.5 \mum\
image to the same angular resolution as the 24~\mum\ MIPSGAL
\citep[survey of the inner Galactic plane using MIPS;][]{car09}
image, i.e. 6\arcsec, and overlaying the two images until a good
match was found. Note that the heavy saturation of the MIPSGAL image
does not hinder the comparison between it and the smoothed COMICS
image. In practice, the emission in MIPSGAL is basically pointlike
and the MIPS PSF presents a well defined circular structure even far
from the (saturated) peak. The match between the smoothed COMICS
image and the MIPSGAL image can thus be performed successfully on
such circular features. Inspection by eye indicates a positional
error of less than 1\arcsec, comparable to the astrometric accuracy
of MIPSGAL.

\section{RESULTS AND ANALYSIS}

This section reports our main results from the observation of the
radio continuum, individual maser species and infrared emission
towards the HMSFR \Ga. In the following, the term ``spot'' is used
to refer to maser emission in a single velocity channel, whereas the
term ``feature'' indicates a collection of spots emitting at a
similar position over contiguous channels (i.e. an individual
masing-cloud). In this work, the parameters of maser emission
(position, intensity, proper motion, etc..) are derived for maser
features.

\subsection{1.3~cm and 3.6~cm Continuum}
\label{cm_res}

\begin{figure}
\includegraphics[angle=-90,scale=0.6]{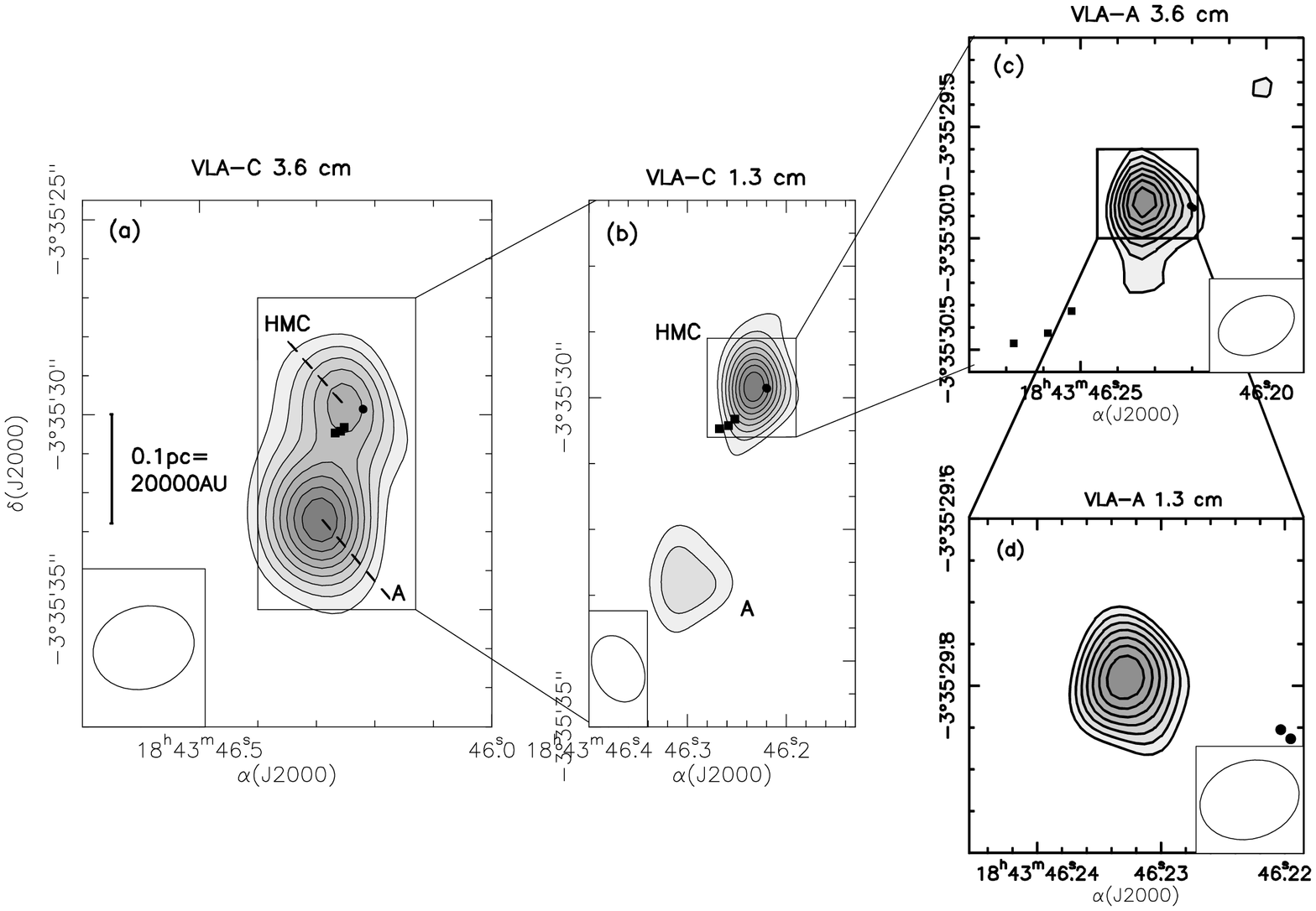}
\caption{Maps (gray scale) of the VLA continuum emission toward \Ga\
at 1.3 and 3.6~cm. (a)~3.6~cm emission with the VLA C-array. \
(b)~1.3~cm emission with the VLA C-array \ (c)~3.6~cm emission with
the VLA A-array \ (d)~1.3~cm emission with the VLA A-array. Squares
and circles report the positions of the 1.665~GHz OH and the 6.7~GHz
\meth\ masers, respectively, detected in this paper (see
Section.~\ref{meth_res}~and~\ref{oh_res}). In each panel, the
contour levels range from 30\% to 90\% of the peak emission at
multiples of 10\%. The restoring beam is shown in the lower, left or
right corner of each panel. A rectangle overlaid on the image
indicates the field of view expanded in the adjacent panel. The
associated labels indicate the names of the continuum sources
identified by us (see Section~\ref{cm_res} and
Table~\ref{tbl:vla-cm}). \label{G28-VLA-cm}}
\end{figure}

Our VLA measurements of the 1.3 and 3.6~cm continuum emission
towards \Ga\ (in both the A-~and~C-array configurations) are
summarized in Table~\ref{tbl:vla-cm}, and shown in
Figure~\ref{G28-VLA-cm}. At the angular resolution of the VLA
C-array, the emission consists of two components, both detected at
1.3~and~3.6~cm. The northern and southern components, separated by
$\sim$3\farcs1, are labeled, respectively, ``HMC'' and ``A'' in
Table~\ref{tbl:vla-cm}. The former name is due to the source being
associated with the hot molecular core imaged by \citet{fur08}.
Source ``A'' is partly resolved with the C-array at 1.3~cm, and is
completely resolved out with the A-array at both 1.3~and~3.6~cm.

In the following, we focus our attention on the centimeter continuum
source ``HMC'', because this appears to be spatially associated with
the observed 22~GHz \wat, 6.7~GHz \meth, and 1.665~GHz OH masers in
\Ga\ (see Figures~\ref{G28-VLA-cm}~and~\ref{G28-absp-l}). While at
1.3~cm the most extended VLA A-array configuration partly resolves
the emission of ``HMC'' (missing about half of the flux measured
with the C-array), at 3.6~cm the emission is essentially compact in
both configurations (yielding comparable integral fluxes). Note
however that the VLA A-array image at 3.6~cm shows a spur protruding
to the south. The peak positions of the 1.3~and~3.6~cm emissions
coincide within the uncertainties. The absolute position of the
A-array 1.3~cm image has been obtained by aligning positions of
persistent water maser spots between the VLA and VLBA observations,
and should be accurate to $\approx 0\farcs1$.

Figure~\ref{G28-hmc-SED-cm} shows the spectral energy distribution
(SED) at centimeter wavelengths of ``HMC''. A spectral index
($\alpha$) of 0.6 is derived from a linear fit of the VLA C-array
1.3~and~3.6~cm and B-array 6~cm flux densities. The latter flux is
an upper limit obtained from the GPS 6~cm Epoch~2 images
downloadable from the MAGPIS web
site\footnote{http://third.ucllnl.org/gps}, which result to be the
most sensitive among the other GPS maps and the relevant map from
the CORNISH survey~\footnote{available on
http://www.ast.leeds.ac.uk/Cornish/public} \citep[see][]{pur08}. We
also fit the SED at centimeter wavelengths with a model of a
homogeneous \ion{H}{2} region, with a Lyman continuum rate of
$7.9\times10^{45}$~s$^{-1}$ and a radius of 0\farcs04.

We caution that the shape of the SED could be significantly affected
by the different sensitivity to extended structures at different
wavelengths. For the same VLA configuration, the shorter the
wavelength, the larger can be the fraction of emission resolved out
by the interferometer. Therefore, one should consider the
possibility that the slope between 3.6~cm and 1.3~cm is steeper than
0.6.

\subsection{Masers Results}

\subsubsection{22.2~GHz Water Maser}
\label{wat_res}

\citet{for99} observed the HMSFR \Ga\ with the VLA C-array (HPBW of
2\farcs2 $\times$ 1\farcs6) and identified 13 water maser spots,
mainly grouped in two clusters separated by $\sim$0\farcs2 in the
northeast-southwest direction. The absolute position of the
brightest spot (with a flux density of 61.3~Jy at the V$_{\rm LSR}$
of 106.3~\kms) differs by $\sim$0\farcs1 from that of the
corresponding spot in our VLBA observation. The field of view of our
VLBA observations covers the whole area within which water maser
emission was detected by \citet{for99}.

Figure~\ref{G28-maser-spe} (upper panel) shows the total-power
spectrum of the 22.2~GHz masers toward \Ga. This profile was
obtained by averaging the total-power spectra of all VLBA antennae,
after weighting each spectrum with the antenna system temperature
($T_{\rm sys}$). Water maser absolute positions and LSR velocities
are plotted in Figure~\ref{G28-absp-l}. Imaging the range of LSR
velocities from 93.2 to 114.3~\kms, we detected 168 distinct water
maser features distributed over a region of about \ 0\farcs3
$\times$ 0\farcs5. Most (92\%) of them concentrate in a small area
of about \ 0\farcs3 $\times$ 0\farcs2 centered on the VLA 1.3~cm
continuum peak. The line-of-sight (l.o.s.) velocity distribution is
remarkably bipolar, with red-shifted features located to the west of
the radio continuum and blue-shifted ones to the east. While the
blue-shifted features mainly concentrate in two small (size
$<$20~mas) clusters, the distribution of red-shifted masers consists
of one narrow strip elongated east-west (size of $\approx$100~mas),
together with two diffuse ``clouds'' of features spread to the north
and to the south of the densely populated, narrow strip. It is also
interesting to note the presence of two streamlines of water
features (enclosed in dashed boxes in Figure~\ref{G28-absp-l} upper
panel), oriented north-south and extending over a few tenths of an
arcsecond, on either side of the VLA 1.3~cm continuum. We took
particular care in filtering out spurious emission owing to poor
cleaning of the beam side lobes (elongated mostly along the
north-south direction) and can confirm that these water maser
streamlines are real.

The individual feature properties are presented in
Table~\ref{tbl:water}. The position offsets are relative to
feature \#1. The absolute positions of feature \#1 at the four
observing epochs are given in Table~\ref{tbl:mas-abs-pos}. Feature
intensities range from 0.1 to 54~Jy. The spread in LSR velocities
ranges from 111.99~\kms\ for the most red-shifted feature (\#107) to
93.33~\kms\ for the most blue-shifted one (\#155).

Only 61 features (36\% of the total) persisted over at least 3
epochs, 34 of which lasting 4 epochs. We calculated the geometric
center (hereafter ``center of motion'', identified with label \#0 in
Table~\ref{tbl:water}, and indicated by a star in
Figure~\ref{G28-rel-pro-s}) of features with a stable spatial and
spectral structure, persisting over the 4 observing epochs, and
refer our measurement of proper motions to this point. With a time
baseline of 9 months, relative (sky-projected) velocities of water
maser features are derived with a mean uncertainty of $\sim$23\%.
The magnitude of relative proper motions ranges from \
1.9$\pm$2.8~\kms \ to \ 58.3$\pm$2.6~\kms, with a mean value of \
24.0~\kms. Figure~\ref{G28-rel-pro-s} shows the derived water maser
proper motions. While the two blue-shifted clusters move to the east
and northeast, most velocities of the red-shifted water masers point
either to the south or to the west. In particular, looking at the
distribution of the maser velocity vectors along the (east-west)
narrow strip outlined by the red-shifted features, we note that
maser features closer to the ``center of motion" move mainly to the
south. Getting closer to the western end of the maser strip, the
orientation of the maser velocity vectors changes gradually from
south to west. The general pattern of water maser relative
velocities appears to indicate expansion from a point on the sky
close to the ``center of motion''. Note also that at larger
distances from the ``center of motion'', maser velocities seem to be
larger and better collimated along the (north)east-(south)west
direction. In particular, we identify two H$_2$O maser clusters
located at the northeast and southwest ends of the maser
distribution (named ``J$_b$'' and ``J$_r$'', respectively; see
Figure~\ref{G28-rel-pro-s}), which move with high velocities
directed close to the northeast-southwest direction.

Figure~\ref{G28-abs-pro} (upper panel) reports the intrinsic
absolute proper motions of the water maser features, derived from
the observed absolute proper motion of feature~\#1 corrected for the
apparent motion. The latter has been calculated adding the
contribution of the parallax, the solar motion with respect to the
LSR \citep{sch10}, and the differential Galactic rotation, from a
flat rotation curve with Galactic constants (R$_0 = 8.3\pm0.23$~kpc,
$\Theta_0 = 239\pm7 $~\kms) as recently determined by \citet{bru11}.
The obtained absolute proper motions differ from the relative proper
motions presented in Figure~\ref{G28-rel-pro-s} by a vector of
amplitude $\approx$40~\kms\ pointing to the southwest, which seems
to be the dominant component of all proper motions. We think that
this difference could be caused by the large systematic error due to
the uncertainties in the source distance and solar and Galactic
motion, and/or by a peculiar velocity of the maser source with
respect to its LSR reference system. Our VLBA water maser
observations were not designed to measure the source parallax. With
a positional accuracy of $\approx$0.1--0.2 milli-arcseconds, and
four epochs spanning (only) 9 months, the observed absolute proper
motion of feature~\#1, not corrected for the apparent motion, looks
similar to a straight line. From the small deviations from linear
motion, we derive an upper limit of \ 0.2~mas \ for the amplitude of
the sinusoidal parallax signature. This corresponds to a lower limit
of \ 5~kpc \ for the source distance, in agreement with the adopted
kinematical distance of 7.4~kpc.

We have checked the reliability of the relative, water maser proper
motions by following an alternative approach. As described in
Section.~\ref{meth_res}, we have also derived the absolute proper
motion of the strongest 6.7~GHz methanol maser feature. To date,
VLBI observations of 6.7~GHz \citep{san10a,san10b} and 12~GHz
\citep{mos03,mos10,mos11} methanol masers toward several sources
indicate that these masers usually show small proper motions, with
typical velocities of only a few km~s$^{-1}$. Assuming that also in
\Ga\ methanol masers move significantly slower than the water
masers, we can use the absolute proper motion of the methanol masers
to correct the absolute proper motion of the water masers, and
derive the water maser velocities as approximately seen in a
reference system comoving with the star. Figure~\ref{G28-abs-pro}
(lower panel) shows the absolute water maser proper motions after
subtracting the absolute methanol maser proper motion (evaluated in
Section.~\ref{meth_res}). Note that, within the measurement errors,
the new absolute proper motions are fully consistent with the
relative proper motions calculated with respect to the ``center of
motion'' (Figure~\ref{G28-rel-pro-s}). This result reinforces our
assertion that the ``center of motion'' can define a suitable
reference system to calculate water maser velocities. In the
following discussion, we use the relative velocities of the water
masers to describe the gas kinematics close to the (proto)star.

\subsubsection{6.7~GHz Methanol Maser}
\label{meth_res}

Figure~\ref{G28-maser-spe} (middle panel) shows the total-power
spectrum of the 6.7~GHz methanol masers observed on 2008 March 18
toward \Ga\ using the Effelsberg 100-m telescope. The 6.7~GHz
methanol maser spectrum, with two narrow features at V$_{\rm LSR}$
of about 97 and 105~\kms, resembles that of the water masers
(Figure~\ref{G28-maser-spe}, upper panel). Note that \citet{cas95}
detected only the emission at V$_{\rm LSR}$ of 105~\kms\ with the
Parkes 64-m telescope.

We imaged the range of LSR velocities from 95 to 107~\kms\ with the
EVN and detected two distinct methanol maser features, both with
V$_{\rm LSR}$ close to 105~\kms. Feature \#1 is detected at the
first, third, and fourth epoch, whereas feature \#2 is detected at
the third and fourth epoch only. Due to adverse atmospherical
conditions and telescope failures, no 6.7~GHz maser signal was
detected at the second epoch. Individual feature properties are
presented in Table~\ref{tbl:meth}. Positional offsets are relative
to feature \#1, and the absolute positions of feature \#1 at the
three epochs of detection are given in Table~\ref{tbl:mas-abs-pos}.
The two methanol maser features lie at the western border of the
distribution of the red-shifted water masers, $\sim$0\farcs2 to the
southwest of the VLA 3.6~cm continuum peak (see
Figure~\ref{G28-absp-l}). Inspecting our images, we could not find
any reliable emission at the velocity of $\sim$97~\kms, in
correspondence of the weakest feature detected in the total-power
spectrum. The mean rms noise on the channel maps was
20~mJy~beam$^{-1}$, and we could have missed that emission if
falling below the 5$\sigma$ detection threshold. Because of the
narrow line width ($\le$1~\kms) and the offset in V$_{\rm LSR}$ from
the systemic velocity (101.0~\kms) of the HMC,
 we tend to exclude that the weak methanol feature at \
V$_{\rm LSR}$ $\sim$97~\kms\ is thermal emission from the HMC
resolved out by the EVN observations.

After correcting for the apparent motion, the absolute proper motion
for feature \#1 is: $V_{\alpha}=-27.8\pm1.0$,
$V_{\delta}=-45.7\pm4.6$~\kms. As discussed in Sect.~\ref{wat_res},
we have used this vector to correct the absolute proper motions of
the water masers. The relative motion of the methanol maser feature
\#1 with respect to the ``center of motion'' of the water masers is
found to be: $v_{\alpha}=1.7\pm6.7$, $v_{\delta}=-15.6\pm8.0$~\kms.
This is consistent with the results from other VLBI observations
which show that 6.7~GHz masers move significantly slower than water
masers, and justifies employing the methanol maser absolute motion
to correct the water maser absolute motions.

\subsubsection{1.665~GHz OH Maser}
\label{oh_res}

\citet{for99} observed the HMSFR \Ga\ with the VLA A-B hybrid
configuration (HPBW of \ 5\farcs5 $\times$ 1\farcs2) in the left
circular polarization (LCP) band and detected eight 1.665~GHz OH
maser spots. Their brightest spot had a flux density of 9.85~Jy at
the V$_{\rm LSR}$ of 102.8~\kms.

Figure~\ref{G28-maser-spe} (lower panel) shows the $T_{\rm
sys}$-weighted mean of the VLBA antenna total-power spectra of the
1.665~GHz OH maser emission in the LCP and RCP bands. Three strong,
narrow features are visible in the LCP band over the LSR velocity
range from 102 to 108~\kms. Two weak features are detected in the
RCP band, one of them at an LSR velocity of $\sim$108~\kms. These
findings are similar to the results obtained by \citet{cas83} with
the Parkes 64-m telescope.

We imaged the range of LSR velocities from 102 to 110~\kms, and
identified 3 distinct OH maser features in the LCP band. In the RCP
band, by imaging maser data after direct phase-referencing to the
calibrator J1834$-$0301, no signal was detected above a 5$\sigma$
level of 0.1~Jy~beam$^{-1}$. Individual feature properties are
presented in Table~\ref{tbl:oh}. The positional offsets are relative
to feature \#1, whose absolute position is given in
Table~\ref{tbl:mas-abs-pos}. The derived absolute position is offset
by $\sim$0\farcs9 from the reference position of \citet{for99}.
Maser intensities range from 0.7 to 1.9~Jy~beam$^{-1}$. The LSR
velocities vary from 102.8~\kms\ for the most blue-shifted feature
(\#1), to 105.5~\kms\ for the most red-shifted feature (\#3). The
FWHM line width of individual maser features ranges from 0.4 to
1.4~\kms. Positions and LSR velocities of hydroxy1 maser features
are plotted in Figure~\ref{G28-absp-l}. The three detected features
lie $\approx0\farcs6$ to the southeast of the VLA 3.6~cm continuum
source and are distributed along a line extended $\approx 0\farcs3$.

\subsection{24.5~\mum\ Infrared Emission}

Figure~\ref{G28-q-cm} shows an image of the 24.5~\mum\ continuum emission
obtained with Subaru/COMICS at an angular resolution of 0\farcs75.
One can see that the structure of the emitting region is quite complex.
Beside two components associated with the radio sources ``HMC'' and ``A''
with a total flux of $\sim$80~Jy, another fainter source ($\sim$3~Jy) is
visible to the south, offset by $\sim$4\arcsec\ from ``A''.

The emission from the northern region is clearly resolved and
appears to be roughly associated with the two radio objects,
although one cannot rule out the existence of additional, fainter
unresolved sources contributing to the extended halo around the two
main sources. In particular, the 24.5~\mum\ emission from ``HMC''
seems to split into two sub-components in the E-W direction, with
comparable flux densities: $\sim$32~Jy for the eastern and
$\sim$43~Jy for the western component. These could trace two
distinct objects or diffuse emission surrounding a single source at
the origin of the ``HMC'' radio continuum. A third, more intriguing
possibility is that one is looking at an ``hour-glass'' shaped
nebulosity centered on the embedded source at the origin of the
free-free emission.

In the latter scenario, the source lies at the center of a flattened
core lying approximately along the line of sight. This dusty core
absorbs most of the IR photons emitted at 24~$\mu$m towards the
observer, while the IR radiation can escape along the minor axis of
the core, lying close to the plane of the sky. We have estimated an
optical depth of $\sim 4$ for the HMC at 24~$\mu$m, assuming the
H$_2$ column density of $4.2\times10^{23}$~cm$^{-2}$ obtained by
\citet{fur08} and the opacity adopted by \citet{whi03a} -- see their
Figure~1. Such a value could be sufficiently large to explain the
lack of emission towards the embedded star and determine the
``hour-glass'' shape of the IR emission. Interestingly, the E--W
orientation of this putative bipolar nebula is also roughly
consistent with the direction of the H$_2$O maser jet in
Figure~\ref{G28-rel-pro-s}, suggesting that one could be observing a
bipolar flow piercing through a dense core, and revealed on
different scales in different tracers. However, such a scenario,
albeit appealing, is not consistent with the inclination of the
maser jet with respect to the plane of the sky. In fact, the
blue-shifted emission lies to the E (see Figure~\ref{G28-rel-pro-s})
and this implies that the eastern lobe should be pointing towards
the observer and thus be brighter than the western lobe in the IR.
Looking at Figure~\ref{G28-q-cm} one can see that this is not the
case, being the emission slightly stronger to the west (see above).

In conclusion, we consider both the ``hour-glass'' nebula and the
double source hypotheses too speculative on the basis of the
observational evidence collected so far, and in the following we
make the conservative assumption that the IR emission is associated
with ``HMC'', with a flux of $\sim$75~Jy.

The southern source ``A'' has a 24.5~\mum\ flux density of
$\sim$5~Jy. Also in this case, the association between the IR
emission and the radio source is questionable, because the two are
offset each other by $\sim$1\arcsec. However, this is the
astrometrical error of the IR image and we thus believe that such an
offset is not sufficient evidence against a common origin for the IR
and radio emission toward ``A''. Therefore, we conclude that both
emissions can originate from the same object.

\subsection{Spectral Energy Distribution}
\label{sed}

Figure~\ref{G28-SED-IR} shows the SED of \Ga\ from 3.6~\mum\ to
1.1~mm. Beside our COMICS data, we have used data from the following
surveys: Galactic Legacy Infrared Mid-Plane Survey Extraordinaire
\citep[GLIMPSE;][]{ben03}, MIPSGAL \citep{car09}, MSX \citep{ega03},
IRAS \citep{neu84}, the Herschel infrared Galactic Plane Survey
\citep[Hi-GAL;][]{mol10a, mol10b}, the APEX Telescope Large Area
Survey of the Galaxy \citep[ATLASGAL;][]{sch09}, and the Bolocam
Galactic Plane Survey \citep[BGPS;][]{dro08}. For the
Herschel/Hi-GAL data, the most recent release (obtained in July 2011
with HIPE version 7.0.0) was used, resulting from data reduction
with the RomaGal software \citep{tra11}, where the most relevant
image artifacts have been removed.

Apart from the IRAS and MSX flux densities, which have been taken
from the corresponding Point Source Catalogues (sources
IRAS\,18411-0338 and G028.8621+00.0657), all the other values have
been calculated by integrating the emission inside the dashed
contour in Figure~\ref{inf-ima}. The latter was chosen in such a way
to include the whole emitting region at all wavelengths. In this way
we overcome the problem of different angular resolutions at
different wavelengths. In the calculation of the flux the background
emission was subtracted assuming that the background flux inside the
contour is in all points equal to the typical flux density measured
along the contour itself. The flux densities used for the SED are
given in Table~\ref{tbl:tsed}.

Figure~\ref{inf-ima}
illustrates the pc-scale structure of the \Ga\ star forming region,
imaged with Herschel from 70 to 500~\mum\ and with APEX at 870~\mum,
and demonstrates that from the mid-IR to the sub-mm the continuum
emission is dominated by a compact source surrounded by a weaker
halo.

We have fitted the SED with the radiative transfer model developed
by \cite{rob07} and \cite{whi03a, whi03b} (using the SED fitting
tool available on http://caravan.astro.wisc.edu/protostars/), which
assumes a ZAMS star with a circumstellar disk, embedded in an
infalling envelope, and allows derivation of a number of physical
parameters. While we are aware that this model cannot adequately
reproduce a complex situation such as that of the multiple MYSOs in
\Ga, we believe that the fit is sufficiently good to guarantee good
estimates of {\it integrated} quantities such as the luminosity and
the mass of the envelope. The best-fit values are, respectively,
2.3$\times$10$^{5}~L_\sun$, and 3.0$\times$10$^{3}~M_\sun$. Note
that in the fitting procedure, the IRAS fluxes were considered upper
limits due to the limited angular resolution, whereas the MIPSGAL
24~\mum\ flux is in fact a lower limit because the corresponding
image is saturated and only the non saturated pixels have been
considered in the calculation. Also, the distance was kept fixed to
the nominal value of 7.4~kpc. The best-fit value for the visual
foreground extinction is 46~mag.

\section{DISCUSSION}

\subsection{Nature of the Radio Continuum}
\label{dis_1}

\begin{figure}
\includegraphics[angle=-90,scale=0.6]{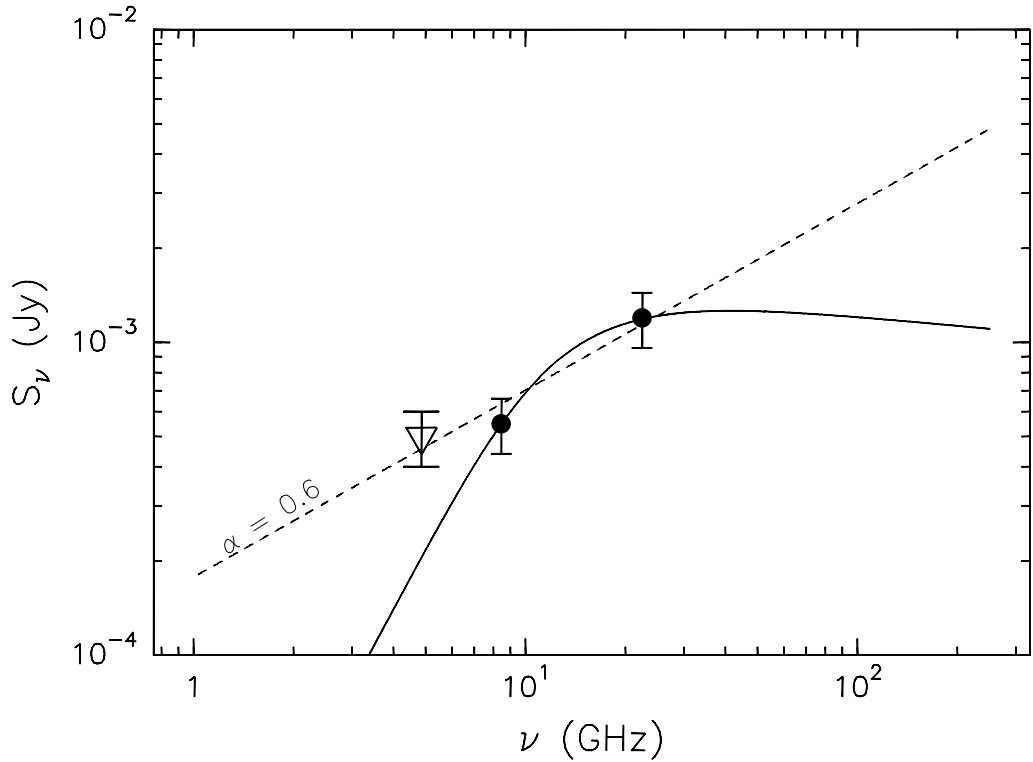}
\caption{Spectral energy distribution of the VLA component ``HMC''
toward the HMSFR \Ga. Dots and error bars report the values and the
associated errors (20\%) of the fluxes at 1.3 and 3.6~cm measured
with the VLA C-array (see Table~\ref{tbl:vla-cm}). The triangle
indicates the upper limit of the flux at 6~cm obtained with the VLA
B-array \citep{pur08}. The linear fit to these measurements,
indicated by the dotted line, produces a spectral index $\alpha =
0.6$. The solid line shows the best fit of the measured fluxes with
a model of a homogeneous \ion{H}{2}~region, yielding a Lyman
continuum rate of $7.9 \times 10^{45}$~s$^{-1}$ \ and a
\ion{H}{2}~region radius of 0\farcs04. \label{G28-hmc-SED-cm}}
\end{figure}

\begin{figure}
\includegraphics[scale=0.6]{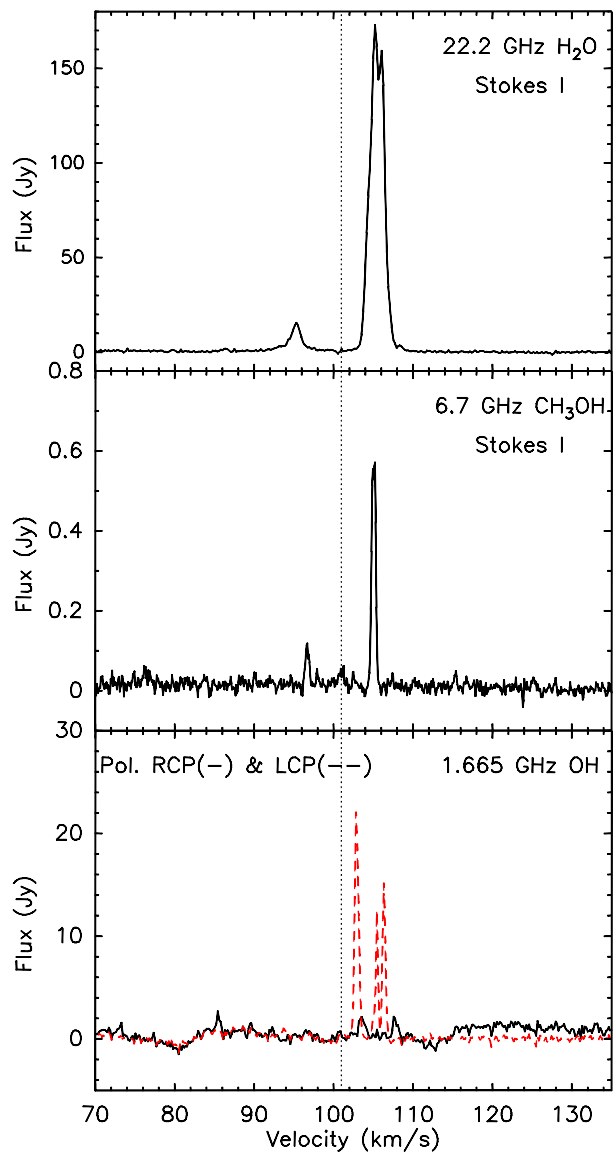}
\caption{\scriptsize Total-power spectra of the \wat, \meth, and OH
masers toward \Ga. Upper panel:~system-temperature (T$_{\rm sys}$)
weighted average of the 22.2~GHz total-power spectra of the 9 VLBA
antennas observing on 2006 April 23. \ Middle panel:~Effelsberg
total-power spectrum of the 6.7~GHz methanol maser emission on 2008
March 18. \ Lower panel:~T$_{\rm sys}$-weighted average of the
1.665~GHz total-power spectra of the 9 VLBA antennas observing on
2007 April 28. Continuous (black) and dashed (red) lines are used to
distinguish between right (RCP) and left (LCP) circular
polarizations, respectively. The dotted line crossing the spectra
denotes the systemic velocity (V$_{\rm sys}$) inferred from NH$_3$
measurements. \label{G28-maser-spe}}
\end{figure}

\begin{figure}
\includegraphics[angle=-90,scale=0.5]{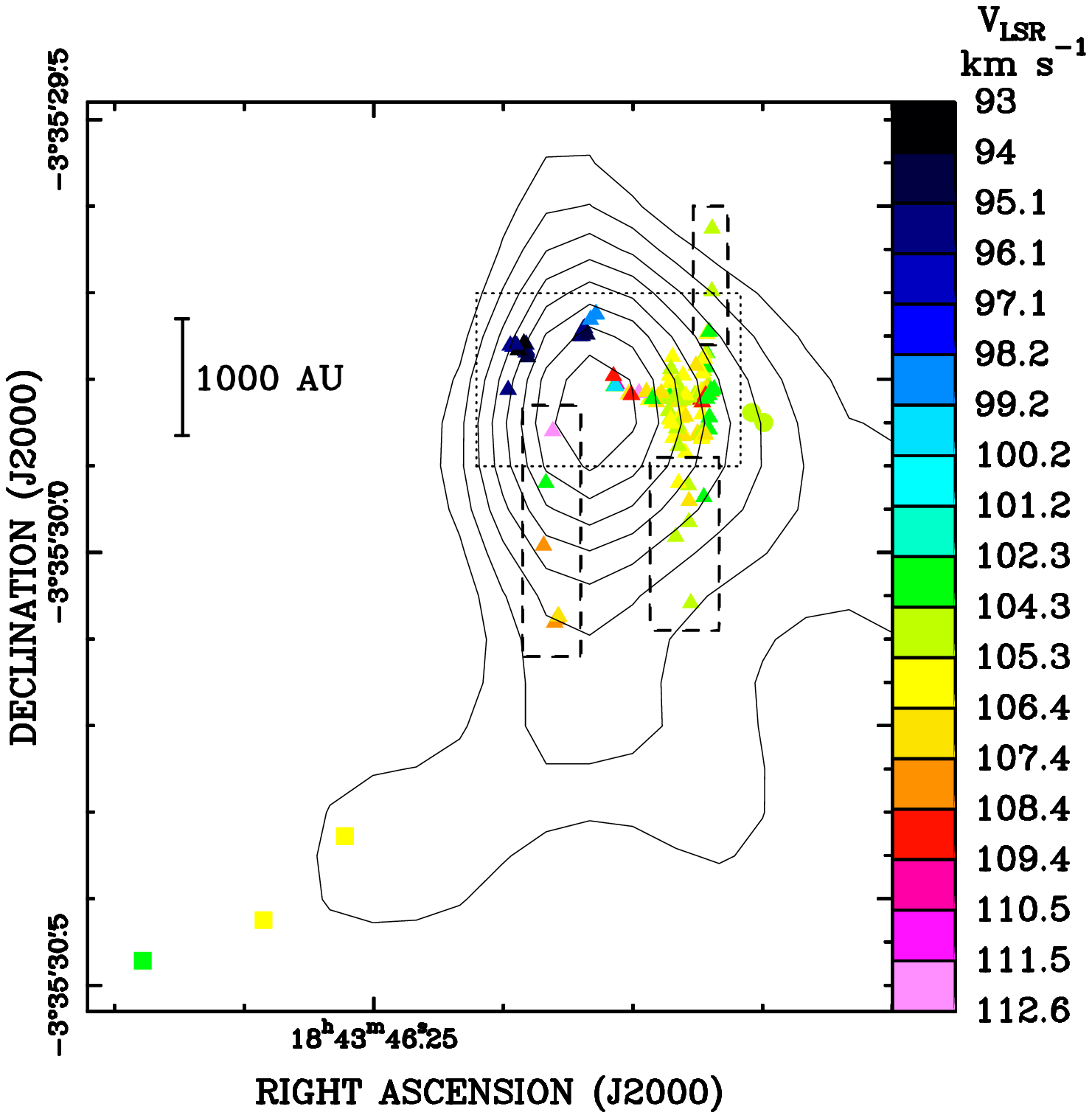}\\
\includegraphics[angle=-90,scale=0.45]{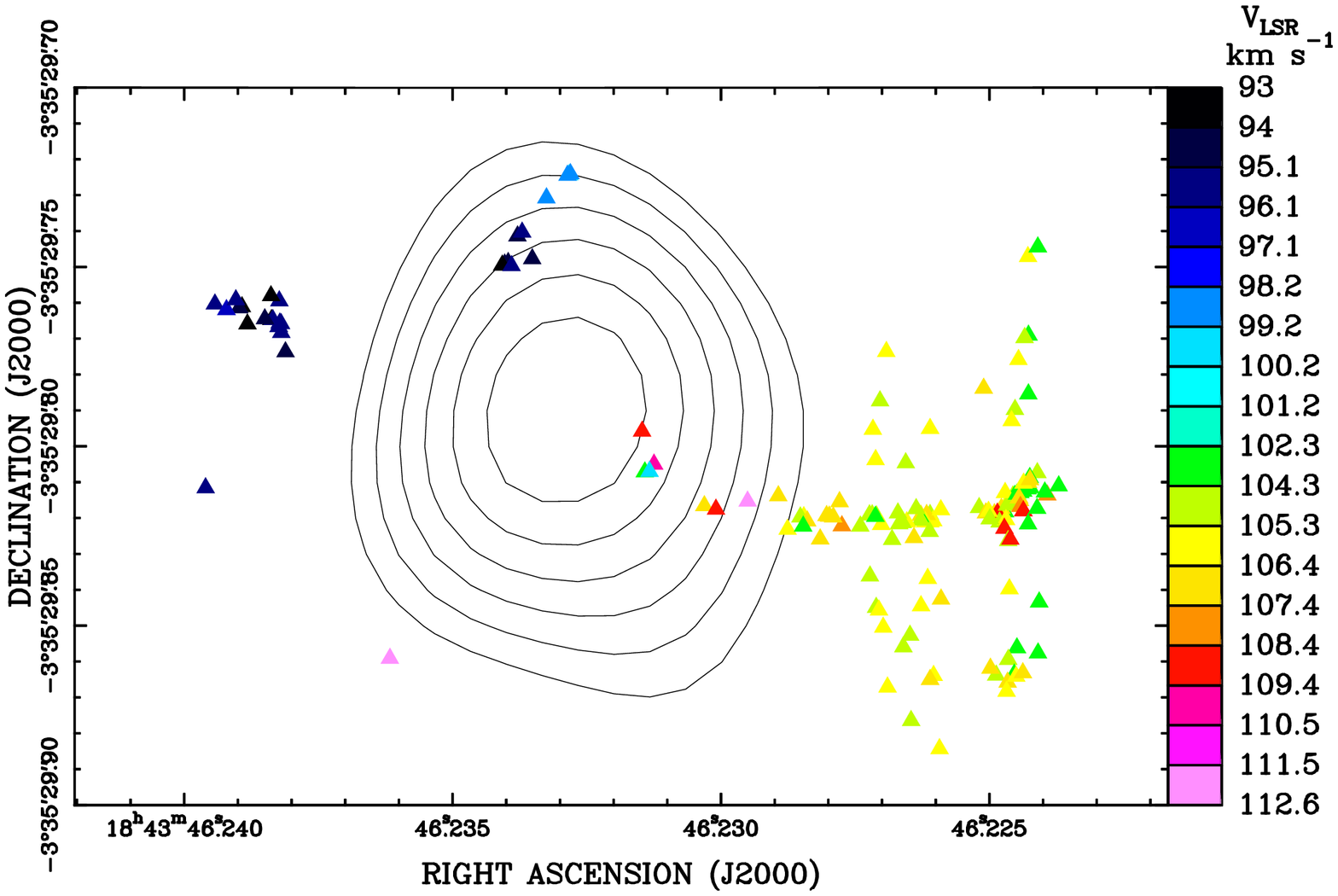}
\caption{\scriptsize Spatial distribution of 22~GHz \wat\
(triangles), 6.7~GHz \meth\ (dots), and 1.665~GHz OH (squares)
masers in \Ga. Different colors are used to indicate the maser LSR
velocities, according to the color scale on the righthand side of
the plot. Upper Panel:~global view of the region where the three
maser species distribute. The solid contours give the 3.6~cm
continuum emission observed with the VLA A-array. Contour levels
range from 20\% to 90\% of the peak emission (0.35~mJy beam$^{-1}$)
at multiples of 10\%. Dashed rectangles enclose the two north-south
streamlines of water masers observed on either side of the VLA
3.6~cm continuum. A dotted rectangle overlaid on the image indicates
the field of view expanded in the right panel. Lower Panel:~zoom on
the area where most of water masers concentrate. The solid contours
give the 1.3~cm continuum emission observed with the VLA A-array.
Contour levels range from 40\% to 90\% of the peak emission
(0.62~mJy beam$^{-1}$) at multiples of 10\%. \label{G28-absp-l}}
\end{figure}

\begin{figure}
\includegraphics[angle=-90,scale=0.54]{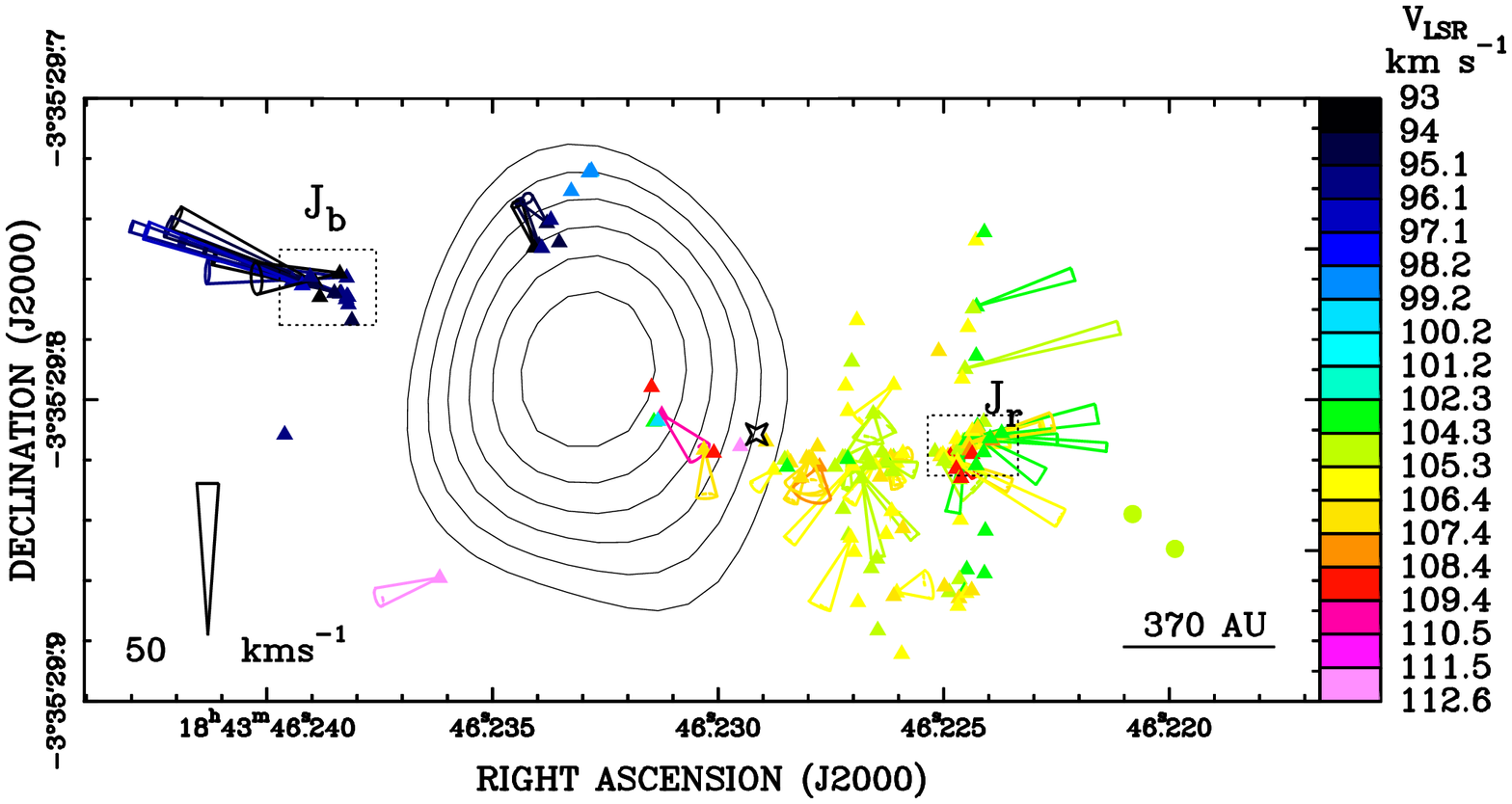}\\
\includegraphics[angle=-90,scale=0.44]{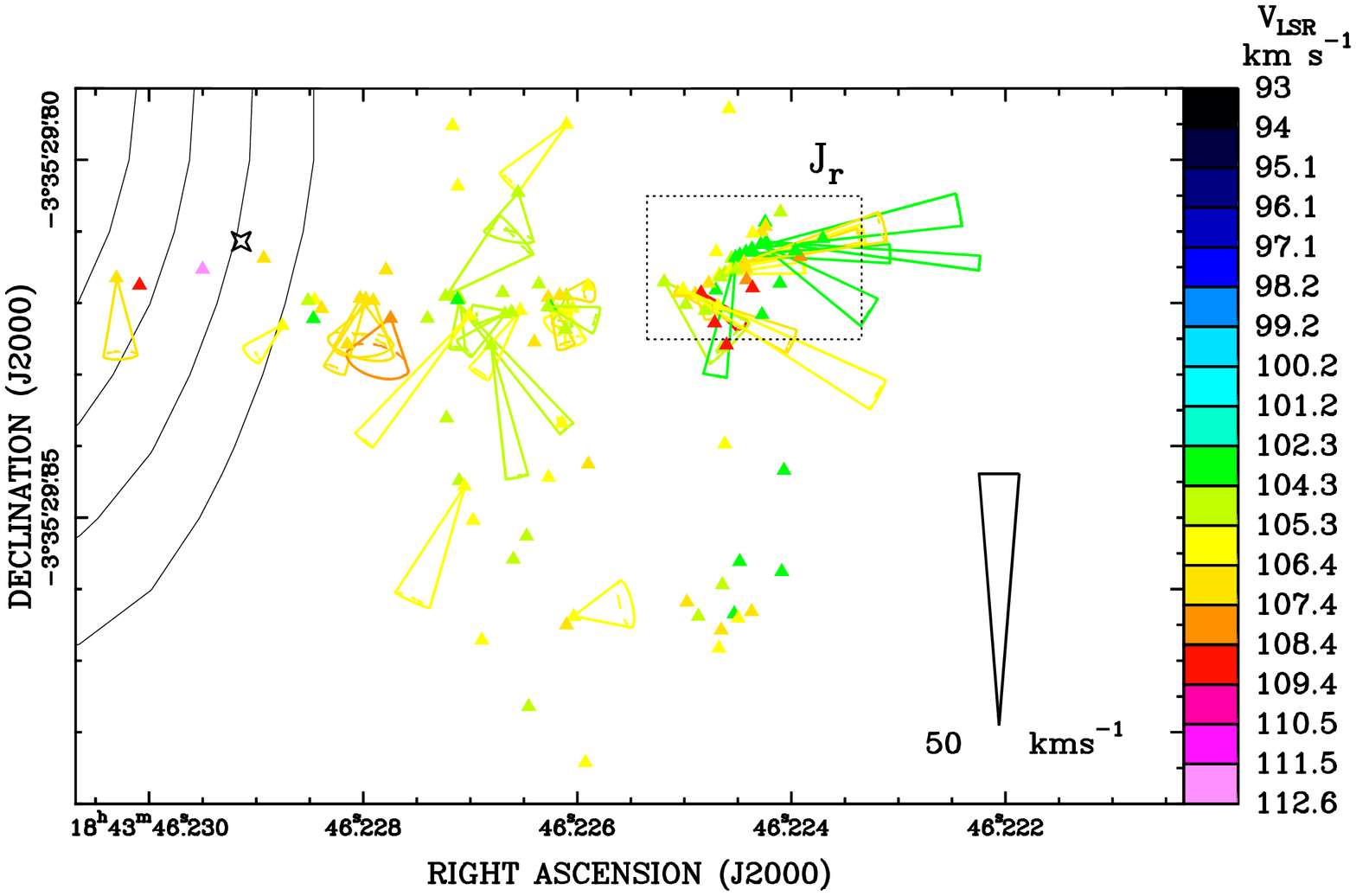}
\caption{\scriptsize Proper motions of 22~GHz \wat\ masers
(triangles) relative to their ``center of motion", and spatial
distribution of 6.7~GHz \meth\ masers (dots) in \Ga. Different
colors are used to indicate the maser LSR velocities, according to
the color scale on the righthand side of the plot. The cones
indicate the 3-D velocities of the water maser features relative to
the ``center of motion'' (marked with a cross). Points without an
associated cone have been detected only over one or two epochs and
the associated proper motion cannot be computed or is considered
unreliable. The cone opening angle gives the 1$\sigma$ uncertainty
on the proper motion direction. The length of the cone is
proportional to the velocity, with the amplitude scale indicated in
the lower left corner of the figure. Open dotted rectangles mark two
clusters of water masers located at the northeast and southwest edge
of the distribution (labeled ``J$_b$'' and ``J$_r$'', respectively),
moving fast and close to the northeast-southwest direction. The VLA
1.3~cm continuum emission is plotted with solid contours. Contour
levels range from 40\% to 90\% of the peak emission (0.62~mJy
beam$^{-1}$) at multiples of 10\%. Upper Panel:~view of the whole
distribution of water maser proper motions. Lower Panel:~zoom on the
most crowded area of water maser concentration, to the west of the
VLA 1.3~cm continuum.\label{G28-rel-pro-s}}
\end{figure}

\begin{figure}
\includegraphics[angle=-90,scale=0.7]{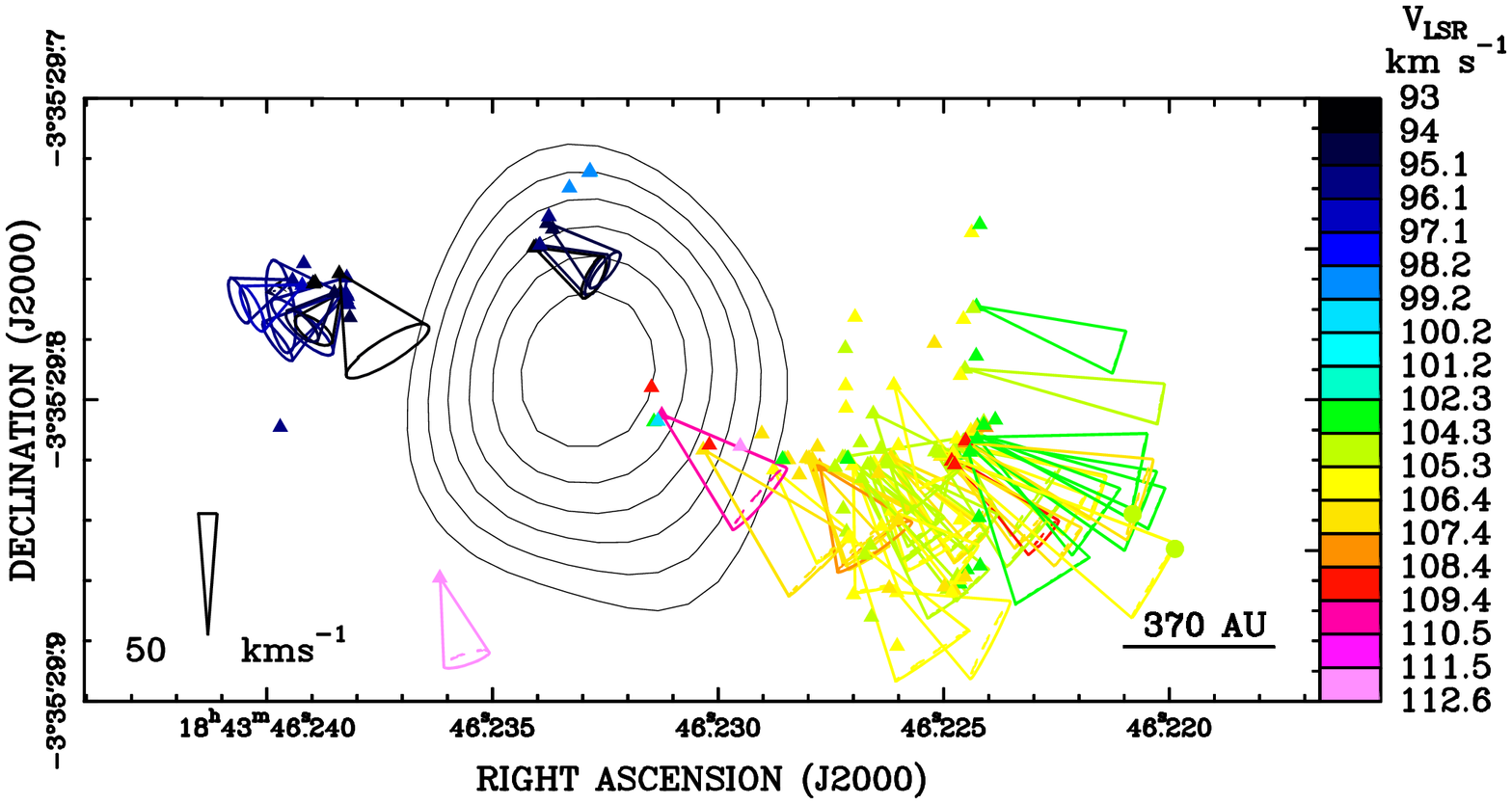}\\
\includegraphics[angle=-90,scale=0.7]{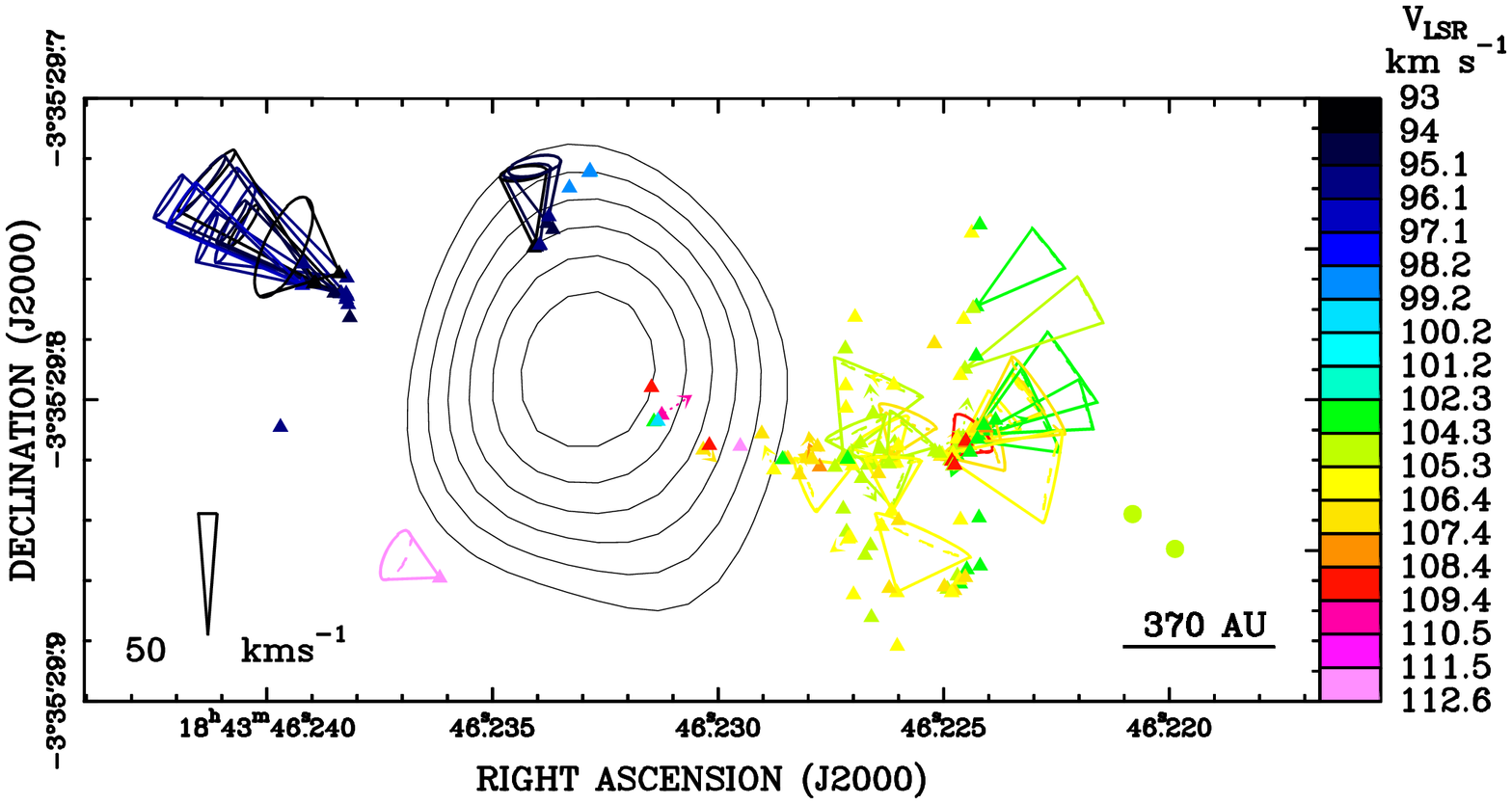}
\caption{\scriptsize Absolute proper motions of water masers in \Ga,
derived by applying different corrections. Upper Panel:~Correcting
for the apparent motion due to the combination of the parallax, the
solar motion with respect to the LSR, and the differential Galactic
rotation (adopting the flat rotation model indicated in
Section~\ref{mas_obs}). Lower Panel:~Correcting for the absolute
proper motion of feature~\#1 of the \meth\ masers. Symbols, colors,
and contour levels have the same meaning as in
Figure~\ref{G28-rel-pro-s}. The cones indicate the 3-D velocities of
the water maser features. Points without an associated cone have
been detected only over one or two epochs and the associated proper
motion cannot be computed or is considered unreliable. The cone
opening angle gives the 1$\sigma$ uncertainty on the proper motion
direction. The length of the cone is proportional to the velocity,
with the amplitude scale indicated in the lower left corner of the
figure.\label{G28-abs-pro}}
\end{figure}

\begin{figure}
\includegraphics[scale=0.45]{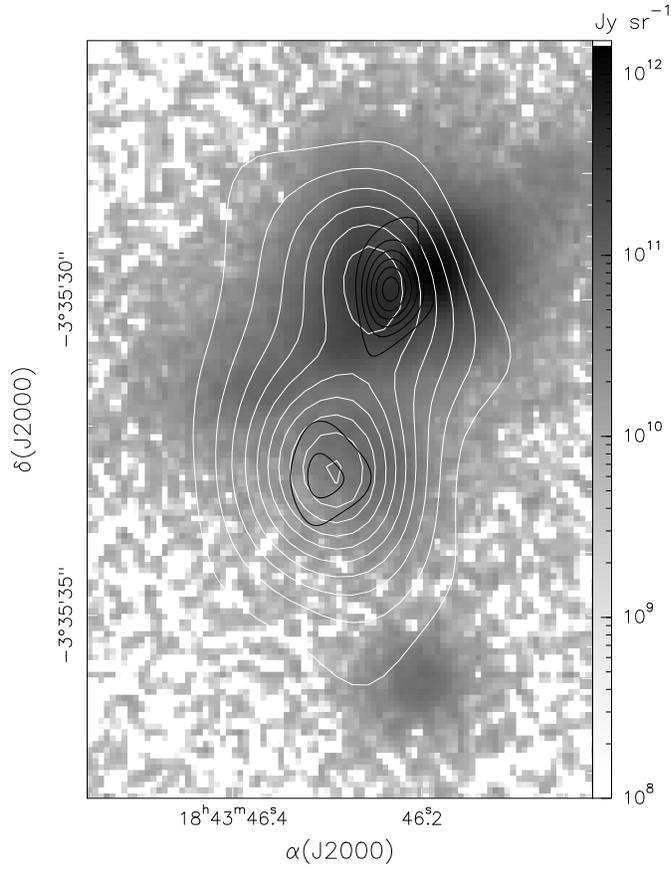}
\caption{Continuum emission maps toward \Ga\ at centimeter and
infrared wavelengths. The grey image in the background is the
24.5~\mum\ emission from Subaru/COMICS. The white and black contours
indicate the centimeter continuum emission at 3.6 and 1.3~cm,
respectively. The values of the contours are the same as in
Figure~\ref{G28-VLA-cm}. \label{G28-q-cm}}
\end{figure}

In this section, we focus on the cm-continuum source labeled ``HMC''
in Table~\ref{tbl:vla-cm}, found in good correspondence with the
positions of the 22~GHz \wat, 6.7~GHz \meth, and 1.665~GHz OH masers
in \Ga\ (see Figure~\ref{G28-absp-l}). The continuum spectrum (shown
in Figure~\ref{G28-hmc-SED-cm}) is consistent with free-free
emission from an ultracompact~(UC)~\ion{H}{2}~region, with the
turnover frequency in the range 10--20~GHz, and ionized by a Lyman
continuum rate of 7.9$\times$10$^{45}$~s$^{-1}$. This would
correspond to a ZAMS star of spectral type B1--B0.5 \citep{pan73},
with a luminosity of \ $L_{star} \sim 8\times10^{3}$~$L_{\odot}$,
and a mass of \ $M_{star} \sim 10~M_{\odot}$. It is worth mentioning
that these values would be different if more recent studies than the
seminal article by Panagia were adopted \citep[e.g.,][]{cro05}.
However, the basic result that one is dealing with an early B-type
star with a luminosity of $\sim10^4~L_\odot$ would remain valid.

We note that the fit to the continuum spectrum requires a very small
size of the UC~\ion{H}{2} region, of $\sim$0\farcs04 or
$\sim$300~AU, which appears in contradiction with the fact that our
A-array 1.3~cm observations partly resolve the continuum emission
(see Table~\ref{tbl:vla-cm}).

Since we considered a homogeneous distribution of ionized gas in our
fit, one possibility is that the \ion{H}{2} region is not
homogeneous. The southern spur identified in the A-array 3.6~cm map
of the ``HMC'' source (see Figure~\ref{G28-absp-l}), could indeed
suggest the existence of a density gradient toward the south. One
may speculate that being source ``A'' more extended and perhaps more
evolved, the gas in its surroundings is less dense. In this
scenario, the north-south streamlines of water masers on either
sides of the southern spur, might trace shocks at the interface
between the ionized gas of ``HMC'' (expanding to the south) and the
surrounding molecular gas.

We also note that the continuum spectrum of ``HMC'' between 1.3 and
6~cm can be also fitted with a power law $S_\nu\propto\nu^\alpha$,
with $\alpha = 0.6$ (see Figure~\ref{G28-hmc-SED-cm}). This spectral
index is typical of thermal jets \citep{rey86} and provides us with
an alternative explanation for the origin of the free-free emission.
However, a caveat is in order. As discussed in Section~\ref{cm_res},
the slope of the spectrum between 3.6~cm and 1.3~cm could be
significantly affected by the inteferometer resolving out part of
the emission at the shorter wavelength. If the spectrum is steeper
than estimated from our data, the slope could be still consistent
with a thermal jet, but also with that of an \ion{H}{2} region with
a density gradient \citep[see e.g.,][]{fra00}.

We conclude that it is impossible to decide the nature of the free-free
emission on the basis of the information collected so far. Notwithstanding
this caveat, as a working hypothesis we prefer the jet interpretation,
because this permits to interpret the continuum and maser emission in the
same context, as discussed in the following section.

\subsection{The Water Maser Flow}
\label{dis_2}

In this section we propose that both the continuum and the water
maser emission are associated with a thermal jet impinging against
the surrounding quiescent molecular gas. This idea is based on two
observational findings. First of all, Figure~\ref{G28-rel-pro-s}
shows that, at larger distances from the center of the water maser
distribution, maser (relative) velocities appear to be larger and
better collimated along the northeast-southwest direction. Secondly,
the axis of the powerful, molecular outflow mapped at arcsecond
scales toward \Ga\ has a northeast-southwest orientation
\citep[Figure~5]{fur08}. This it is plausible to assume that the
two, fast moving, water maser clusters ``J$_b$'' and ``J$_r$'',
located, respectively, at the northeast and southwest edge of the
maser distribution, are excited by the same jet responsible also for
driving the large-scale molecular outflow. One could object against
the ``jet" interpretation, that the whole pattern of water maser
velocities cannot be explained only in terms of a collimated
outflow. In fact, at smaller distances from the center of the maser
distribution, water maser velocities point to the north or south
directions, and form large angles with the direction of the putative
jet. We propose that the scattered pattern of proper motions
presented by the red-shifted water masers can result from the
interaction of the jet with dense circumstellar material, which
deflects and increase the opening angle of the flow velocities.

If the free-free continuum source ``HMC'' is a jet, the position of
the (proto)star powering it may be offset from the free-free
continuum peak. If a jet powered by the (proto)star drives the water
maser motions, and, in particular, the expansion to the south and
west of the narrow strip of red-shifted water masers, the
(proto)star could be located at an intermediate position between
``HMC'' and the red-shifted maser strip. This point is not far
(within tens of mas) from the position of the center of motion of
the water masers (cross in Figure~\ref{G28-rel-pro-s}). Considering
the large uncertainties in deriving the center of water maser
expansion, in the following discussion we can safely take the center
of motion as a reasonable estimate of the (proto)star location.

In the jet scenario, one can estimate the momentum rate of the jet
from the average distance of the H$_2$O masers from the MYSO and the
average maser velocity. Assuming a H$_{2}$ pre-shock density of
10$^{8}$~cm$^{-3}$, as predicted by models \citep{eli89}, the
momentum rate in the water maser jet is given by the expression
$\dot{P} = 1.5 \times
10^{-3}V_{10}^{2}R_{100}^{2}(\Omega/4\pi)$~$M_{\odot}$~yr$^{-1}$~km~s$^{-1}$.
Here $V_{10}$ is the average maser velocity in units of
10~km~s$^{-1}$, $R_{100}$ the average distance of water masers from
the YSO in units of 100~AU, and $\Omega$ the solid angle of the jet.
This expression has been obtained by multiplying the momentum rate
per unit surface transferred to the ambient gas ($n_{\rm H_{2}}\,
m_{\rm H_{2}}\, V^{2}$), by $\Omega R^{2}$, under the assumption
that the jet is emitted from a source at a distance $R$ from the
masers within a beaming angle $\Omega$. To estimate the average
distance and velocity of the water masers tracing the jet in \Ga, we
use all the maser features belonging to the two fast moving clusters
``J$_b$'' and ``J$_r$'' (see Figure~\ref{G28-rel-pro-s}). The
resulting average distance is 122~mas (or 902~AU) and the average
speed is 43~km~s$^{-1}$. Using these values, the jet momentum rate
is $\dot{P} = 2.2 \,
(\Omega/4\pi)$~$M_{\odot}$~yr$^{-1}$~km~s$^{-1}$.

For shock induced ionization in a thermal jet, the momentum rate of
the jet can be estimated from the continuum flux assuming the
emission is optically thin (see \cite{ang96}). One has: $F_\nu d^{2}
= 10^{3.5}(\Omega/4\pi)\dot{P}$, where $F_\nu$ is the measured
continuum flux density in mJy, $\dot{P}$ the jet momentum rate in
$M_{\odot}$~yr$^{-1}$~km~s$^{-1}$, $\Omega$ the jet solid angle in
sterad, and $d$ the source distance in kpc. Using the flux of
1.2~mJy measured at 1.3~cm with the VLA C-array for source ``HMC'',
and the distance of 7.4~kpc, one derives $\dot{P} = 2 \times 10^{-2}
\,(\Omega/4\pi)^{-1}$~$M_{\odot}$~yr$^{-1}$~km~s$^{-1}$.

Note that the momentum rates estimated from the maser motion and the
continuum flux have a different dependency on the jet solid angle,
which can thus be estimated by equating the two momentum rate estimates.

We obtain an expression for the jet solid angle: \ $ 2.2 \,
(\Omega/4\pi) = 2 \times 10^{-2} \, (\Omega/4\pi)^{-1}$. We find $
\Omega = 1.2$~sterad, i.e. a jet semi-opening angle \ $\theta_j =
35$\degr. Correspondingly, the jet momentum rate is \ $\dot{P} =
0.2~M_{\odot}$~yr$^{-1}$~km~s$^{-1}$. This value is large enough to
justify the momentum rate of several
$10^{-2}~M_{\odot}$~yr$^{-1}$~km~s$^{-1}$, estimated by
\citet{fur08} for the large-scale molecular outflow. This result
supports our hypothesis that the small-scale thermal and H$_2$O
maser jet is sufficiently powerful to feed the large-scale molecular
outflow observed by \citet{fur08}. We note that the derived
parameters for the jet in \Ga\ are comparable
 to those of the few collimated outflows observed up to now
in massive star-forming regions.  For comparison, the jets in
HH~80--81 \citep{mar98}, in  IRAS~16547$-$4247 \citep{rod08}, and in
IRAS~20126+4104 \citep{mos11} have momentum rates of the order of \
$10^{-1}~M_{\odot}$~yr$^{-1}$~km~s$^{-1}$ and opening angles of \
20--30\degr.

\subsection{Methanol Maser Environment}
\label{dis_3}

The two detected 6.7~GHz methanol maser features are approximately
aligned along the same northeast-southwest line defined by the two
maser jet clusters ``J$_b$'' and ``J$_r$'', the peak of the VLA
A-array 1.3~cm, and the center of motion (see
Figure~\ref{G28-absp-l}). Figure~\ref{G28-maser-spe} shows that the
total-power spectra of the 6.7~GHz methanol and 22~GHz water masers
are similar, both presenting a strong red-shifted and a much weaker
blue-shifted component. The positions of the methanol maser features
roughly along the jet axis, and the similarities in the spectral
shape of the water and methanol maser emissions, make us speculate
that also the 6.7~GHz methanol masers could originate in the
(proto)stellar jet, albeit associated with more quiescent gas than
that traced by the water masers. Maser VLBI observations toward the
sources IRAS\,20126$+$4104 and \Gc\ \citep{mos11,san10b} have also
indicated that the 6.7~GHz masers might trace entrained gas at the
interface with a jet.

Figure~\ref{G28-absp-l} clearly shows that the blue-shifted lobe of
the water maser jet is sampled much more sparsely, as it presents
much less features than the red-shifted lobe. This is consistent
with the blue-shifted line being significantly weaker than the
red-shifted one in the total power spectrum (see
Figure~\ref{G28-maser-spe}, upper panel). This fact suggests that
the circumstellar material is not isotropically distributed around
the (proto)star, but it is mostly concentrated on the side of the
red-shifted lobe. That would also account for the stronger emission
line in the total-power spectrum of the \meth\ masers at the
red-shifted velocities.

\subsection{OH Maser Environment}
\label{bozomath}

The spectrum of the 1.665~GHz maser is dominated by three narrow LCP
features (see Figure~\ref{G28-maser-spe}, lower panel), and is very
similar to the spectrum of the 1.667~GHz maser transition observed
by \cite{cas83}. These authors find that most OH interstellar masers
are highly polarized, either RCP or LCP, with the same predominant
polarization for both the 1.665~GHz and 1.667~GHz transition. They
propose an explanation based on large-scale velocity and/or magnetic
field gradients across the masing region.

Figure~\ref{G28-VLA-cm} shows that the 1.665~GHz OH maser features
are found significantly offset ($\approx0\farcs6$) from the compact
VLA source ``HMC'' and the center of water maser activity, and are
closer to the more extended source ``A''. A possible interpretation
is that OH masers trace more external, less dense parts of the
molecular core surrounding the (proto)star responsible for the
excitation of the water and methanol masers. Observations indicate
that the density of massive proto-stellar cores decreases from the
center outward, with a power-law index of about 1.6 ($n \propto
r^{-1.6}$) \citep{beu02a}. In \Ga, the ratio of the average distance
(from the ``center of motion'' of water masers) of the water
($\approx0\farcs1$) and OH ($\approx0\farcs6$) masers, is about 6,
implying a decrease in gas density by a factor of about 20. That
agrees with maser excitation models, which predict a H$_{2}$
pre-shock density of \ 10$^{6}$--10$^{7}$~cm$^{-3}$ for OH 1.665~GHz
\citep{ces91,cra02}, and \ 10$^{8}$~cm$^{-3}$ for \wat\ 22~GHz
masers \citep{eli89}, respectively.

\subsection{The Nature of Source ``HMC''}

So far, we have explored the properties of source ``HMC'' making use
of the maser, radio, and IR continuum emission. Now, we want to use
our findings to shed light on the nature of the MYSO(s) associated
with this source.

We have seen that the radio continuum emission and the water maser
features suggest the presence of a jet with a large momentum rate,
on the order of $\sim10^{-1}~M_\odot$~km~s$^{-1}$~yr$^{-1}$. Values
that high are consistent with the jet being powered by a high-mass
(proto)stars, as indicated, e.g., by the relationship between
outflow momentum rate and MYSO luminosity obtained by \cite{beu02b}.
In particular, from their Figure~4b, one can conclude that the MYSO
must have a luminosity of at least several $10^3~L_\odot$.
Alternatively, if the radio emission were originating from an
\ion{H}{2} region, we have seen (Section~\ref{dis_1}) that the
ionizing star should have a luminosity $\la10^4~L_\odot$. Clearly,
both the thermal jet and the \ion{H}{2} region hypotheses imply
similar luminosity for the associated MYSO.  The question we want to
investigate here is whether such a luminosity is consistent with the
results obtained from the IR data.

In Section~\ref{sed} we have found that the total luminosity of the
region is on the order of $2\times10^5~L_\odot$. However, this
includes all MYSOs in a radius $\ga$10\arcsec\ around ``HMC'',
whereas we need to estimate the contribution of the sole ``HMC'', to
achieve a consistent comparison with the luminosity estimate
obtained from the water maser and/or radio continuum. Our 24.5~\mum\
continuum image can be used for this purpose. As shown in
Figure~\ref{G28-q-cm} , the ``HMC'' source is dominating the
emission at this wavelength, contributing to $\sim$90\% of the total
flux density, and a similar conclusion holds also for the GLIMPSE
images. Due to the lack of sub-arcsecond imaging at longer
wavelengths, it is difficult to establish whether also the flux
density measured close to the peak of the SED is mostly due to
source ``HMC''. Nonetheless, the Hi-GAL/Herschel data suggest that
the 70~$\mu$m emission peaks closer to source ``HMC'' than source
``A''. A word of caution is in order here. The angular resolution of
the Hi-GAL data at 70~\mum\ is estimated to be about
$10\arcsec\times9\arcsec$ (after reduction with the Roma-Gal
software), while the astrometrical error is $\sim$3\farcs5 (the
latter results from aligning the PACS 70~$\mu$m images to the
MIPSGAL 24~$\mu$m images by matching suitable point sources detected
at both wavelengths). Although large, these values still permit to
associate the 70~\mum\ emission with source ``HMC''. In fact, the
angular separation between the 3.6~cm continuum source ``HMC'' and
the 70~$\mu$m peak is 3\farcs7, consistent with the positional
error, whereas source ``A'' lies 6\farcs5 from the 70~$\mu$m peak.
This can be appreciated in Figure~\ref{fcm70}, where one sees that
the 70~$\mu$m emission peaks to the NW of ``HMC''. Therefore, in the
following we assume that the bolometric luminosity estimated from
the SED is likely dominated by the ``HMC'' source and conclude that
the MYSO(s) embedded in the HMC may be as luminous as
$\sim2\times10^5~L_\odot$.

Notwithstanding the large uncertainties of all our luminosity estimates, the
value obtained from the IR images exceeds by an order of magnitude
that derived from the maser and radio continuum data. This
inconsistency can be explained if ``HMC'' is hosting a multiple
stellar system rather than a single star. One can get an idea of how
this may affect our estimate of the stellar properties (mass, etc.)
by assuming that the luminosity of $2\times10^5~L_\odot$ is emitted
by a stellar cluster with a \cite{mil79} initial mass function. One
finds that the luminosity of the most massive star may be as small
as $\sim3\times10^4~L_\odot$, depending on the statistical
distribution of the stellar masses in the cluster. Given the
uncertainties, this value is only marginally greater than the
luminosity estimate from the radio continuum and maser data.

The same method provides us with the total mass of the stellar
cluster, which ranges from $5\times10^2~M_\odot$ to
$2\times10^3~M_\odot$, depending on the statistical distribution of
the stars in the cluster. This stellar mass can be divided by the
total mass of the associated molecular clump, $3\times10^3~M_\odot$
(see Section~\ref{sed}), to calculate the star formation efficiency,
17\%--67\%. Albeit large, values like these are not implausible, as
we are considering the part of the molecular cloud where star
formation concentrates \citep{lad03, bon11}.

In conclusion, we believe that the observational evidence collected
so far for source ``HMC'' suggests that one is observing a young
star of 20--30~$M_\odot$, still in the main accretion phase and
surrounded by a rich stellar cluster. Such a conclusion holds
independently of the scenario adopted to explain the free-free
emission (thermal jet or weak \ion{H}{2} region) and raises a
question: why only faint free-free emission is detected from this
object? A $\sim3\times10^4~L_\odot$ star is expected to ionize the
surrounding gas emitting a flux density of $\sim$2~mJy (in the
optically thin limit), whereas towards source ``HMC'' less than
1~mJy is measured. A possible explanation is that the \ion{H}{2}
region is at least partly quenched by accretion from the parental
core, as depicted, e.g., in the model by \citet{ket02, ket03}.
Although no evidence of accretion is seen in a typical infall tracer
such as HCO$^+$ \citep{fur08}, this fact is inconclusive because the
majority of the detected HCO$^{+}$(1--0) might be part of the
outflowing gas, thus hindering the detection of infall.

Finally, it is also worth mentioning that according to recent
theoretical models \citep{hos10}, large accretion rates are bound to
bloat the protostellar radius and decrease the corresponding
temperature and Lyman continuum flux. This could be an alternative
explanation of the lack of a (strong) \ion{H}{2} region.

\section{SUMMARY}

We observed the HMSFR \Ga\ with the VLBI phase-referencing technique
in three powerful maser transitions: 22.2~GHz \wat, 6.7~GHz \meth,
and 1.665~GHz OH. In addition, we also performed VLA observations of
the radio continuum emission at 1.3~and~3.6~cm with both the
A-~and~C-array, and Subaru/COMICS observations of the mid-infrared
continuum emission. We also made use of data from the
Hi-GAL/Herschel, ATLASGAL/APEX, as well as other surveys of the
Galactic plane. From all these observations, we draw the following
conclusions:

\begin{enumerate}

\item
The bipolar distribution of l.o.s.
velocities and the general pattern of observed relative proper
motions indicate that the water masers are tracing expansion with an
average velocity of $\sim$24~\kms\ from a point on the sky close to
the maser ``center of motion''. While at larger distance from the
``center of motion'', maser velocities are higher and better
collimated along the northeast-southwest direction, at closer
distances water masers show smaller speeds and a larger scatter in
the velocity orientations. We interpret these facts as
the result of the interaction of a (proto)stellar jet with dense circumstellar material.

\item Two continuum sources (named ``HMC'' and ``A''),
separated by $\sim$3\farcs1 along the north-south direction, are
observed with the VLA at 3.6~and~1.3~cm. The ``HMC'' source is
compact and spatially associated with the observed \wat, \meth\ and
OH masers. Whether the free-free emission arising from this source
is due to an \ion{H}{2} region or a thermal ionized jet cannot be
unambiguously established with the present data. However, we believe
that the jet hypothesis is more consistent with the direction of the
water maser motions, which outline expansion in a bipolar flow.
Moreover, the orientation and momentum rate of the putative jet are
compatible with those of the molecular outflow detected at arcsecond
scales. We hence argue that the jet traced by the water masers is
also powering the large-scale outflow.

\item The combined information obtained from sub-arcsecond imaging with
Subaru/COMICS at 24.5~\mum\ and lower angular resolution images from
the Hi-GAL/Herschel and ATLASGAL/APEX surveys, permit to estimate
with good accuracy the luminosity ($2\times10^5~L_\odot$) and gas
mass ($3\times10^3~M_\odot$) of the star forming region of \Ga, and
establish that $\sim$90\% of the luminosity is coming from the radio
source ``HMC''. We conclude that this source must contain multiple
stars, with the most massive being at least as luminous as
$3\times10^4~L_\odot$. The lack of an associated H{\sc ii} region
indicates that the 20--30~$M_\odot$ star must be still undergoing
heavy accretion from the surrounding envelope.

\end{enumerate}

\acknowledgments

We thank the anonymous referee for careful reading of the text and the
constructive criticisms that improved the presentation and
analysis of our results.
This work was supported by the Chinese NSF through grants NSF
11133008, NSF 11073054, NSF 10733030, and NSF 10703010.


\clearpage

\begin{figure}
\includegraphics[scale=0.7]{g28_sedfit.eps}
\caption{Spectral energy distribution of the VLA component ``HMC''
toward \Ga\ from 3.6~\mum\ to 1.1~mm. Dots indicate real values,
whereas triangles with the vertex up and down denote lower and upper
limits, respectively. Section~\ref{sed} lists the infrared
telescopes used to reconstruct the SED. We have fitted the SED with
the radiative transfer model developed by \citet{rob07} (using the
SED fitting tool available on
http://caravan.astro.wisc.edu/protostars/). \label{G28-SED-IR}}
\end{figure}

\begin{figure}
\includegraphics[scale=0.65,angle=-90]{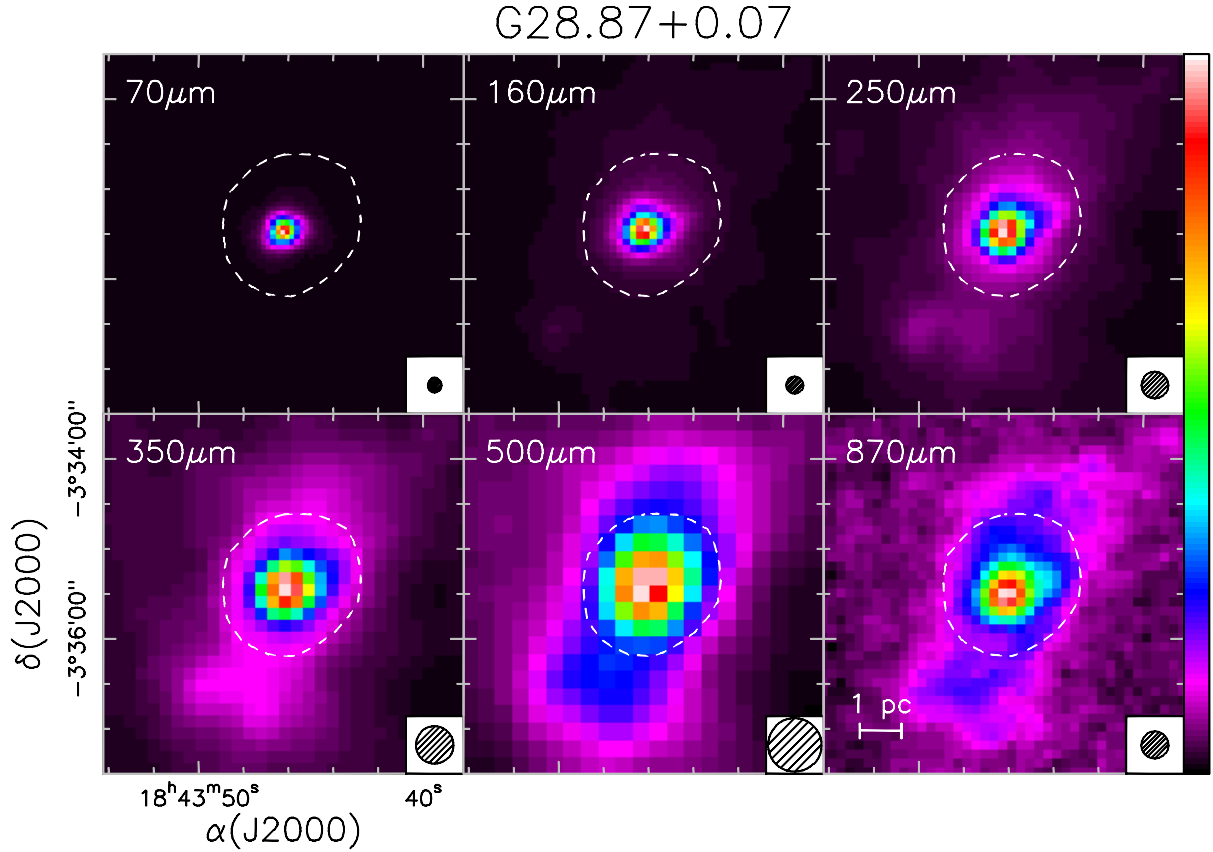}
\caption{Each panel shows the pc-scale structure of the \Ga\
star-forming region at a different infrared wavelength (indicated in
the upper left corner of the panel). Images from 70 to 500~\mum\ are
observed with Herschel, and the map at 870~\mum\ is taken with APEX.
The color scale corresponds to increasing intensity from dark purple
to white, as indicated in the wedge to the right. The observing beam
is reported in the lower right corner of each panel. The dashed
white circle denotes the area of integration to calculate the \Ga\
flux at different infrared wavelengths.\label{inf-ima}}
\end{figure}

\begin{figure}
\includegraphics[scale=0.6,angle=-90]{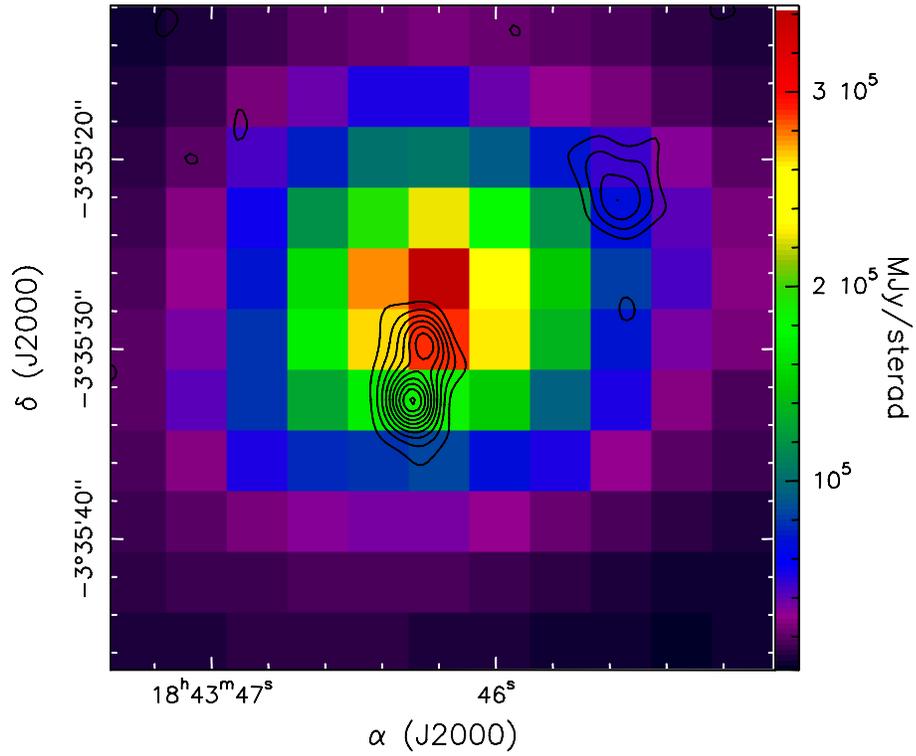}
\caption{Overlay of the 3.6~cm continuum map (contours; same as in
Figure~\ref{G28-q-cm}) on the Hi-GAL/Herschel image at 70~$\mu$m.
Note how the far-IR emission peaks close to the free-free source
``HMC''.\label{fcm70}}
\end{figure}

\begin{deluxetable}{cccccccccc}
\tabletypesize{\scriptsize} \tablecaption{Radio continuum emission
in \Ga.\label{tbl:vla-cm}} \tablewidth{0pt} \tablehead {\colhead{} &
\colhead{} & \colhead{} & \colhead{} & \colhead{} & \colhead{} &
\multicolumn{2}{c}{Peak position} & \colhead{} & \colhead{}\\
\colhead{Label} & \colhead{Array} & \colhead{$\lambda$} &
\colhead{HPBW} & \colhead{PA} & \colhead{Image rms} &
\colhead{R.A.(J2000)} & \colhead{Dec.(J2000)} & \colhead{F$_{\rm
peak}$} & \colhead{F$_{\rm int}$} \\ \colhead{} & \colhead{} &
\colhead{(cm)} & \colhead{($\arcsec \times \arcsec$)} &
\colhead{(\degr)} & \colhead{(mJy beam$^{-1}$)} &
\colhead{$\mathrm{(^h\;\;\;^m\;\;\;^s)}$} &
\colhead{$(\degr\;\;\;\arcmin\;\;\;\arcsec)$} & \colhead{(mJy
beam$^{-1}$)} & \colhead{(mJy)}\\} \startdata
HMC & VLA$-$A & 1.3 & 0.122 $\times$ 0.093 & $-$17 & 0.06 & 18 34 46.23 & $-$03 35 29.8 & 0.62 & 0.64 \\
    & VLA$-$C & 1.3 & 1.053 $\times$ 0.755 & $-$9  & 0.05 & 18 43 46.23 & $-$03 35 29.8 & 1.03 & 1.20 \\
    & VLA$-$A & 3.6 & 0.33  $\times$ 0.24  & $-$1  & 0.03 & 18 43 46.23 & $-$03 35 29.9 & 0.35 & 0.42 \\
    & VLA$-$C & 3.6 & 2.62  $\times$ 2.11  & 5     & 0.02 & 18 43 46.25 & $-$03 35 29.8 & 0.53 & 0.55 \\
A   & VLA$-$A & 1.3 & 0.122 $\times$ 0.093 & $-$17 & 0.06 & \nodata & \nodata & $<$0.18(3$\sigma$) & \nodata \\
    & VLA$-$C & 1.3 & 1.053 $\times$ 0.755 & $-$9  & 0.05 & 18 43 46.30 & $-$03 35 32.8 & 0.39 & 0.60  \\
    & VLA$-$A & 3.6 & 0.33  $\times$ 0.24  & $-$1  & 0.03 & \nodata & \nodata & $<$0.09(3$\sigma$) & \nodata \\
    & VLA$-$C & 3.6 & 2.62  $\times$ 2.11  & 5     & 0.02 & 18 43 46.29 & $-$03 35 32.7 & 0.81 & 0.98 \\
\enddata
\tablecomments{Column~1 gives the source label; columns~2~and~3
report the VLA array configuration and the observing band,
respectively; columns~4~and~5 the (major and minor) HPBW size and
the PA of the observing beam, respectively; column~6 the image rms;
columns~7, 8, 9, and 10 the absolute position, the peak intensity,
and the integral intensity of the emission peak, respectively. The
PA of the beam is defined  East of North.}
\end{deluxetable}

\begin{deluxetable}{lclllcll}
\tabletypesize{\scriptsize}\tablecaption{Positions and
Brightness.}\tablewidth{0pt}\tablehead{ \colhead{Maser} & Feature &
epoch & \colhead{R.A. (J2000)} & \colhead{Dec. (J2000)} &
\colhead{F$_{\rm peak}$} & \colhead{NW beam} \\
\colhead{} & \colhead{} & \colhead{} &
\colhead{$\mathrm{(^h\;\;\;^m\;\;\;^s)}$} &
\colhead{$(\degr\;\;\;\arcmin\;\;\;\arcsec)$} & (Jy beam$^{-1}$) &
\colhead{(mas$\times$mas @ deg)} } \startdata

H$_{2}$O & 1 & 2006 Apr 23 & 18~43~46.22467 & $-$03~35~29.8161  & 57.4 &  1.2$\times$0.4 @ $-$17  \\
         &   & 2006 Jun 30 & 18~43~46.22465 & $-$03~35~29.8173  & 50.3 &  1.3$\times$0.4 @ $-$16  \\
         &   & 2006 Sep 28 & 18~43~46.22455 & $-$03~35~29.8190  & 32.0 &  1.4$\times$0.5 @ $-$3 \\
         &   & 2007 Jan 18 & 18~43~46.22448 & $-$03~35~29.8212  & 20.7 &  1.5$\times$1.0 @ 20  \\
CH$_{3}$OH & 1 & 2006 Feb 28 & 18~43~46.22079 & $-$03~35~29.8379 & 0.044 & 9.9$\times$5.6 @ 5\\
           &   & 2007 Mar 18 & \nodata        & \nodata          & \nodata & \nodata \\
           &   & 2008 Mar 18 & 18~43~46.22031 & $-$03~35~29.8521 & 0.246 & 12.7$\times$5.2 @ 45   \\
           &   & 2009 Mar 17 & 18~43~46.22011 & $-$03~35~29.8595 & 0.175 & 12.6$\times$5.9 @ 13 \\
OH & 1 & 2007 Apr 28 & 18~43~46.26771 & $-$03~35~30.4697 & 1.94 & 17.8$\times$10.9 @ 1 & \\
\enddata
\tablecomments {Column~1 reports the maser species; column~2 gives
the label number of the feature, as given in Tables~\ref{tbl:water},
\ref{tbl:meth}, and \ref{tbl:oh}; column~3 lists the observing date;
columns~4~and~5 report the absolute position; column~6 gives the
peak brightness; column~7 reports the (major and minor) HPBW size
and PA of the naturally-weighted (NW) beam. The PA of the beam is
defined  East of North.} \label{tbl:mas-abs-pos}
\end{deluxetable}

\begin{deluxetable}{clllccccc}
\tabletypesize{\scriptsize} \tablecaption{Parameters of VLBA
22.2~GHz H$_{2}$O maser features in \Ga.\label{tbl:water}}
\tablewidth{0pt} \tablehead{
\colhead{Feature} & \colhead{epochs of} & \colhead{V$_{\rm LSR}$} & \colhead{S$_{\nu}$} & \colhead{$\Delta \alpha$} & \colhead{$\Delta \delta$} & \colhead{V$_{\alpha}$} & \colhead{V$_{\delta}$}  \\
\colhead{Number} & \colhead{detection} & \colhead{(km s$^{-1}$)} &
\colhead{(Jy)} & \colhead{(mas)} & \colhead{(mas)} & \colhead{(km
s$^{-1}$)} & \colhead{(km s$^{-1}$)} \\} \startdata
0   & 1,2,3,4  & 103.17 & \nodata & 66.83$\pm$0.05  & 0.005$\pm$0.05        & 0               & 0              \\
1   & 1,2,3,4  & 106.01 & 53.99 &     0           &     0               & $-$16.9$\pm$1.0 &   1.4$\pm$1.0      \\
2   & 1,2,3,4  & 105.07 & 26.07 &  37.44$\pm$0.05 &    $-$3.02$\pm$0.05 & $-$23.0$\pm$1.2 & $-$26.4$\pm$1.2    \\
3   & 1        & 104.36 & 25.38 &  $-$2.88$\pm$0.05 &       2.60$\pm$0.05 &   \nodata &   \nodata              \\
4   & 1,2,3    & 105.27 & 19.92 &  $-$0.85$\pm$0.05 &       0.25$\pm$0.05 &  $-$2.8$\pm$2.2 &   0.2$\pm$2.3    \\
5   & 2,3,4    & 104.27 & 17.62 &  $-$3.29$\pm$0.05 &       2.79$\pm$0.05 & $-$29.8$\pm$1.7 &   0.2$\pm$1.8    \\
\nodata\\
\enddata
\tablecomments{Column~1 gives the feature label number; column~2
lists the observing epochs at which the feature was detected;
columns~3~and~4 report the value of the intensity-weighted LSR
velocity and flux density of the strongest spot, averaged over the
observing epochs; columns~5~and~6 give the position offsets (with
the associated errors) along the R.A. and Dec. axes relative to the
feature \#1, measured at the first epoch of detection;
columns~7~and~8 give the components of the relative proper motion
(with the associated errors) along the R.A. and Dec. axes, measured
with respect to the reference feature \#0. \\
This table is published in its entirety in the electronic edition of
the Astrophysical Journal. A portion is shown here for guidance
regarding its form and content.}
\end{deluxetable}

\begin{deluxetable}{ccccccccc}
\tabletypesize{\scriptsize} \tablecaption{Parameters of EVN 6.7~GHz
CH$_{3}$OH maser features in \Ga.\label{tbl:meth}} \tablehead{
\colhead{Feature} & \colhead{epochs of} & \colhead{V$_{\rm LSR}$} &
\colhead{F$_{\rm peak}$} & \colhead{I} & \colhead{$\Delta\alpha$} &
\colhead{$\Delta\delta$} & \colhead{v$_{\alpha}$}
& \colhead{v$_{\delta}$}\\
\colhead{} & \colhead{detection}& \colhead{(km s$^{-1}$)} &
\colhead{(Jy beam$^{-1}$)} & \colhead{(Jy)} & \colhead{(mas)} &
\colhead{(mas)} & \colhead{(km s$^{-1}$)} & \colhead{(km s$^{-1}$)}}
\startdata
1 & 1,3,4 &  104.88 & 0.155 & 0.190 & 0 & 0 & 1.7$\pm$6.7 & $-$15.6$\pm$8.1 \\
2 & 3,4   &  105.21 & 0.268 & 0.283 & $-$12.41$\pm$0.18 & $-$11.24$\pm$0.18 & \nodata & \nodata \\
\enddata
\tablecomments{Column~1 gives the feature label number; column~2
lists the observing epochs at which the feature was detected;
columns~3~and~4 provide the intensity-weighted LSR velocity and the
flux density of the strongest spot, respectively, averaged over the
observing epochs; columns~5~and~6 give the position offsets (with
the associated errors) along the R.A. and Dec. axes, relative to the
feature~\#1; columns~7~and~8 give the components of the relative
proper motion (with the associated errors) along the R.A. and Dec.
axes, measured with respect to the reference feature~\#0 (the
``center of motion'') of water masers.}
\end{deluxetable}

\begin{deluxetable}{ccccccc}
\tablewidth{0pt} \tablecaption{Parameters of VLBA 1.665~GHz OH maser
features in \Ga.\label{tbl:oh}} \tablehead{
\colhead{Feature} & \colhead{Pol} & \colhead{V$_{\rm LSR}$} & \colhead{F$_{\rm peak}$} & \colhead{I} & \colhead{$\Delta$x} & \colhead{$\Delta$y} \\
\colhead{} & \colhead{} & \colhead{(km s$^{-1}$)} & \colhead{(Jy
beam$^{-1}$)} & \colhead{(Jy)} & \colhead{(mas)} & \colhead{(mas)}}
\startdata
1 & L & 102.9 & 1.94 & 5.26 & 0 & 0  \\
2 & L & 106.1 & 1.44 & 3.52 & $-$233.57$\pm$0.11 & 143.92$\pm$0.13 \\
3 & L & 105.4 & 0.67 & 1.78 & $-$139.46$\pm$0.17 & 46.91$\pm$0.21 \\
\enddata
\tablecomments{Column~1 gives the feature label number; column~2
indicates the circular polarization of the maser emission;
columns~3,~4, and~5 report the intensity-weighted LSR velocity, and
the intensity and the integral flux of the strongest spot,
respectively; columns~6~and~7 list the position offsets (with the
associated errors) along the R.A. and Dec. axes, relative to the
feature~\#1.}
\end{deluxetable}

\begin{deluxetable}{ccc}
\tabletypesize{\scriptsize} \tablecaption{Flux densities measured
towards the \Ga\ region. \label{tbl:tsed}} \tablewidth{0pt}
\tablehead
{\colhead{$\lambda$} & \colhead{$F_\nu$} & \colhead{instrument} \\
\colhead{($\mu$m)} & \colhead{(Jy)} & \colhead{} \\} \startdata
3.6 & 0.9 & Spitzer/GLIMPSE \\
4.5 & 2.1 & Spitzer/GLIMPSE \\
5.8 & 5.3 & Spitzer/GLIMPSE \\
8.0 & 9.4 & Spitzer/GLIMPSE \\
8.3 & 3.3 & MSX \\
12 & 8.6$^a$ & IRAS \\
12.1 & 7.1 & MSX \\
14.7 & 15.3 & MSX \\
21.3 & 34.0 & MSX \\
24 & $>$21 & Spitzer/MIPSGAL \\
24.6 & 87.4 & Subaru/COMICS \\
25 & 105$^a$ & IRAS \\
60 & 1720$^a$ & IRAS \\
70 & 1790 & Herschel/Hi-GAL \\
100 & $<$3180 & IRAS \\
160 & 1350 & Herschel/Hi-GAL \\
250 & 630 & Herschel/Hi-GAL \\
350 & 200 & Herschel/Hi-GAL \\
500 & 53 & Herschel/Hi-GAL \\
870 & 8.8 & APEX/ATLASGAL \\
1100 & 2.5 & CSO/BGPS \\
\enddata

\vspace*{1mm} $^a$~all IRAS fluxes are assumed to be upper limits in
Figure~\ref{G28-SED-IR}
\end{deluxetable}

\clearpage
\section{ONLINE-ONLY MATERIALS}

\begin{deluxetable}{clllccccc}
\tabletypesize{\scriptsize} \tablecaption{Parameters of VLBA
22.2~GHz H$_{2}$O maser features in \Ga. (Table~\ref{tbl:water} in
the print version.)} \tablewidth{0pt} \tablehead{
\colhead{Feature} & \colhead{epochs of} & \colhead{V$_{\rm LSR}$} & \colhead{S$_{\nu}$} & \colhead{$\Delta \alpha$} & \colhead{$\Delta \delta$} & \colhead{V$_{\alpha}$} & \colhead{V$_{\delta}$}  \\
\colhead{Number} & \colhead{detection} & \colhead{(km s$^{-1}$)} &
\colhead{(Jy)} & \colhead{(mas)} & \colhead{(mas)} & \colhead{(km
s$^{-1}$)} & \colhead{(km s$^{-1}$)} \\} \startdata
0   & 1,2,3,4  & 103.17 & \nodata & 66.83$\pm$0.05  & 0.005$\pm$0.05        & 0               & 0                  \\
1   & 1,2,3,4  & 106.01 & 53.99 &     0           &     0               & $-$16.9$\pm$1.0 &   1.4$\pm$1.0      \\
2   & 1,2,3,4  & 105.07 & 26.07 &  37.44$\pm$0.05 &    $-$3.02$\pm$0.05 & $-$23.0$\pm$1.2 & $-$26.4$\pm$1.2    \\
3   & 1        & 104.36 & 25.38 &  $-$2.88$\pm$0.05 &       2.60$\pm$0.05 &   \nodata &   \nodata              \\
4   & 1,2,3    & 105.27 & 19.92 &  $-$0.85$\pm$0.05 &       0.25$\pm$0.05 &  $-$2.8$\pm$2.2 &   0.2$\pm$2.3    \\
5   & 2,3,4    & 104.27 & 17.62 &  $-$3.29$\pm$0.05 &       2.79$\pm$0.05 & $-$29.8$\pm$1.7 &   0.2$\pm$1.8    \\
6   & 2        & 105.06 & 15.08 &   2.36$\pm$0.05 &      $-$3.68$\pm$0.05 &   \nodata &   \nodata              \\
7   & 2        & 105.14 & 14.36 &  14.86$\pm$0.05 &  $-$146.72$\pm$0.05 &   \nodata &   \nodata                \\
8   & 1,2,3,4  & 105.21 & 12.16 &  36.29$\pm$0.05 &    $-$3.33$\pm$0.05 &   0.5$\pm$1.3 & $-$10.0$\pm$1.3      \\
9   & 1,2,3,4  & 105.45 & 11.21 &  $-$1.71$\pm$0.05 &       0.79$\pm$0.05 & $-$26.3$\pm$1.5 &   7.0$\pm$1.5    \\
10  & 1,2,3,4  & 105.25 &  9.40 &  21.35$\pm$0.05 &    $-$2.95$\pm$0.05 &   0.9$\pm$1.5 &  $-$5.7$\pm$1.5      \\
11  & 1        & 103.92 &  9.39 &  $-$3.79$\pm$0.05 &       3.17$\pm$0.05 &   \nodata &   \nodata              \\
12  & 1,2,3,4  & 104.80 &  9.30 &  29.94$\pm$0.05 &      $-$5.50$\pm$0.05 &   8.2$\pm$1.5 &  $-$6.9$\pm$1.6    \\
13  & 1,2,3    & 104.00 &  9.15 &  $-$5.85$\pm$0.05 &     4.22$\pm$0.05 & $-$21.5$\pm$2.2 & $-$13.5$\pm$2.3    \\
14  & 2        & 105.35 &  8.57 &  $-$1.00$\pm$0.05 &    24.53$\pm$0.05 &   \nodata &   \nodata                \\
15  & 1,2,3,4  & 106.34 &  8.22 &  56.56$\pm$0.05 &      $-$3.51$\pm$0.05 & $-$21.2$\pm$1.4 &   3.0$\pm$1.5    \\
16  & 1,2,3,4  & 106.73 &  7.53 &  $-$1.84$\pm$0.05 &       0.91$\pm$0.05 & $-$25.3$\pm$1.1 &   6.4$\pm$1.2    \\
17  & 2        & 105.12 &  7.33 &  3.31 $\pm$0.05 &     $-$46.40$\pm$0.05 &   \nodata &   \nodata              \\
18  & 2,3,4    & 104.94 &  7.23 &  29.24$\pm$0.05 &  $-$165.12$\pm$0.05 &   \nodata &   \nodata                \\
19  & 1,2,3,4  & 105.87 &  7.22 &  34.93$\pm$0.05 &    $-$5.68$\pm$0.05 &  21.1$\pm$1.6 & $-$25.0$\pm$1.6      \\
20  & 2        & 104.70 &  7.02 & $-$11.90$\pm$0.05 &   192.10$\pm$0.05 &   \nodata &   \nodata                \\
21  & 1,2,3,4  & 105.31 &  6.90 &  29.01$\pm$0.05 &      $-$5.21$\pm$0.06 &   7.6$\pm$1.5 &  $-$0.7$\pm$1.6    \\
22  & 1,2      & 104.23 &  6.82 &  36.56$\pm$0.05 &      $-$3.49$\pm$0.05 &   \nodata &   \nodata              \\
23  & 1,2,3,4  & 105.42 &  6.57 &  27.79$\pm$0.05 &      $-$4.98$\pm$0.06 &   8.3$\pm$1.6 & $-$12.6$\pm$1.7    \\
24  & 2        & 105.11 &  6.48 &  $-$0.06$\pm$0.05 &   $-$42.00$\pm$0.05 &   \nodata &   \nodata              \\
25  & 2        & 105.10 &  6.45 &  15.54$\pm$0.05 &  $-$104.30$\pm$0.05 &   \nodata &   \nodata                \\
26  & 1,2,3    &  95.16 &  6.20 & 202.71$\pm$0.05 &      50.48$\pm$0.05 &  55.3$\pm$1.8 &  18.4$\pm$1.9        \\
27  & 1,2,3,4  & 105.50 &  6.16 &  21.32$\pm$0.05 &      21.04$\pm$0.05 &  11.5$\pm$1.9 & $-$11.6$\pm$1.9      \\
28  & 2        & 104.36 &  5.79 &  $-$1.96$\pm$0.05 &     2.10$\pm$0.05 &   \nodata &   \nodata    \\
29  & 1,2,3,4  & 103.91 &  5.57 &  $-$6.62$\pm$0.05 &     4.52$\pm$0.05 & $-$42.5$\pm$1.3 &  $-$4.2$\pm$1.3    \\
30  & 2,3,4    & 103.78 &  5.30 &  $-$2.48$\pm$0.05 &     2.28$\pm$0.05 &   4.2$\pm$1.9 & $-$23.6$\pm$2.0       \\
31  & 1,2,3,4  & 105.13 &  5.28 &  31.91$\pm$0.05 &      $-$9.86$\pm$0.05 &  $-$5.2$\pm$1.8 & $-$26.4$\pm$2.0  \\
32  & 1,2,4    & 104.22 &  5.23 &  $-$2.89$\pm$0.05 &   $-$40.09$\pm$0.05 & $-$50.6$\pm$2.7 &  $-$8.5$\pm$2.7  \\
33  & 1        & 104.35 &  5.23 &  $-$8.60$\pm$0.05 &       8.75$\pm$0.05 &   \nodata &   \nodata  \\
34  & 1,2,3,4  & 106.26 &  5.21 &  21.33$\pm$0.05 &    $-$4.99$\pm$0.05 &  $-$1.6$\pm$1.6 &  $-$6.3$\pm$1.6    \\
35  & 3        & 104.71 &  5.17 &  14.22$\pm$0.05 &  $-$239.02$\pm$0.05 &   \nodata &   \nodata   \\
36  & 1,2,3,4  & 105.46 &  4.86 &  20.48$\pm$0.05 &      $-$4.65$\pm$0.05 &   4.1$\pm$1.7 &  $-$3.7$\pm$1.7   \\
37  & 1,2,3    & 105.15 &  4.78 &  28.10$\pm$0.05 &      11.48$\pm$0.05 &   1.9$\pm$3.1 &  $-$8.7$\pm$3.2     \\
38  & 2        & 105.54 &  4.72 &  36.95$\pm$0.05 &      13.71$\pm$0.05 &   \nodata &   \nodata    \\
39  & 3        & 104.72 &  4.57 & $-$10.01$\pm$0.05 &   121.21$\pm$0.05 &   \nodata &   \nodata    \\
40  & 1,2,3,4  & 108.62 &  4.46 &   2.52$\pm$0.05 &      $-$2.55$\pm$0.05 &  $-$7.9$\pm$1.2 &  $-$6.2$\pm$1.2   \\
41  & 2        & 104.93 &  4.37 &  36.76$\pm$0.05 &     $-$27.47$\pm$0.05 &   \nodata &   \nodata     \\
42  & 1,2      & 104.71 &  4.32 &  $-$5.05$\pm$0.05 &    46.44$\pm$0.05 &   \nodata &   \nodata   \\
43  & 2        & 105.12 &  4.17 &  25.61$\pm$0.05 &     0.01$\pm$0.05 &   \nodata &   \nodata  \\
44  & 1        & 104.23 &  4.06 &  $-$6.09$\pm$0.05 &    30.67$\pm$0.06 &   \nodata &   \nodata   \\
45  & 1,2,3,4  & 106.20 &  4.03 &  61.07$\pm$0.05 &      $-$7.08$\pm$0.05 &   6.2$\pm$1.7 &  $-$5.8$\pm$1.7  \\
46  & 2        & 105.52 &  3.96 &  33.94$\pm$0.05 &      43.85$\pm$0.05 &   \nodata &   \nodata   \\
47  & 1,2,3,4  & 104.67 &  3.95 &  $-$2.36$\pm$0.05 &    26.39$\pm$0.05 & $-$50.8$\pm$1.8 &  13.5$\pm$1.8   \\
48  & 1,2,3    & 104.04 &  3.91 &  $-$4.62$\pm$0.05 &     3.58$\pm$0.05 & $-$41.3$\pm$2.5 &   7.8$\pm$2.7   \\
49  & 1,2,3,4  & 103.85 &  3.84 &  $-$6.13$\pm$0.05 &    47.09$\pm$0.05 & $-$31.9$\pm$1.7 &  10.5$\pm$1.7   \\
50  & 1,2,3,4  & 105.44 &  3.71 &  21.97$\pm$0.05 &     $-$20.76$\pm$0.05 &   1.2$\pm$1.9 &   1.5$\pm$2.1   \\
51  & 1,2      & 105.06 &  3.67 &  38.10$\pm$0.05 &     $-$20.04$\pm$0.05 &   \nodata &   \nodata     \\
52  & 1        & 105.41 &  3.60 &  $-$0.82$\pm$0.05 &   $-$23.65$\pm$0.05 &   \nodata &   \nodata     \\
53  & 2,3,4    & 105.55 &  3.24 &  35.42$\pm$0.05 &     $-$29.60$\pm$0.05 &  10.5$\pm$2.7 & $-$22.8$\pm$2.7  \\
54  & 1,2,3,4  & 106.82 &  3.22 &  23.89$\pm$0.05 &    $-$3.16$\pm$0.05 &  $-$3.5$\pm$1.4 &  $-$9.8$\pm$1.4  \\
55  & 2,3      & 106.46 &  3.13 &  13.73$\pm$0.05 &  $-$123.75$\pm$0.05 &   \nodata &   \nodata   \\
56  & 4        & 105.35 &  3.07 &   3.11$\pm$0.05 &     $-$47.13$\pm$0.06 &   \nodata &   \nodata  \\
57  & 3,4      & 105.95 &  3.02 &  $-$4.34$\pm$0.05 &     5.38$\pm$0.06 &   \nodata &   \nodata   \\
58  & 2,3,4    & 106.37 &  2.93 &   4.53$\pm$0.05 &    $-$2.16$\pm$0.05 & $-$38.8$\pm$2.4 & $-$21.0$\pm$2.4 \\
59  & 3        & 105.39 &  2.90 &  25.72$\pm$0.05 &     $-$25.59$\pm$0.05 &   \nodata &   \nodata   \\
60  & 2        & 105.01 &  2.86 &   0.23$\pm$0.05 &     1.02$\pm$0.05 &   \nodata &   \nodata   \\
61  & 1        & 104.90 &  2.81 &  $-$1.77$\pm$0.05 &       2.26$\pm$0.05 &   \nodata &   \nodata \\
62  & 4        & 105.34 &  2.77 &   2.65$\pm$0.05 &    $-$5.13$\pm$0.06 &   \nodata &   \nodata  \\
63  & 1,2,3,4  & 105.46 &  2.72 &  37.26$\pm$0.05 &    20.82$\pm$0.05 &  $-$9.8$\pm$1.8 & $-$12.8$\pm$1.8 \\
64  & 3        & 105.83 &  2.63 &  35.05$\pm$0.05 &     $-$48.30$\pm$0.06 &   \nodata &   \nodata  \\
65  & 1,2      & 105.30 &  2.51 &  57.47$\pm$0.05 &    $-$3.62$\pm$0.05 &   \nodata &   \nodata  \\
66  & 4        & 103.22 &  2.40 &  $-$5.42$\pm$0.05 &     3.83$\pm$0.05 &   \nodata &   \nodata  \\
67  & 1,2,3    & 105.21 &  2.34 &  38.23$\pm$0.05 &      $-$3.01$\pm$0.05 & $-$14.0$\pm$3.5 &  12.2$\pm$3.7  \\
68  & 1,2,3    & 106.13 &  2.30 &  20.32$\pm$0.05 &     $-$47.78$\pm$0.05 & $-$10.4$\pm$3.1 &   2.4$\pm$3.3  \\
69  & 1,2,3,4  & 106.85 &  2.28 &   3.34$\pm$0.05 &    $-$2.71$\pm$0.05 &  $-$3.2$\pm$1.4 &  $-$3.8$\pm$1.5  \\
70  & 2,3      & 105.04 &  2.26 &  40.16$\pm$0.05 &      $-$5.89$\pm$0.05 &   \nodata &   \nodata    \\
71  & 2        & 106.54 &  2.21 &  26.23$\pm$0.05 &      $-$8.07$\pm$0.05 &   \nodata &   \nodata    \\
72  & 4        & 105.71 &  2.15 &   0.42$\pm$0.06 &     $-$42.97$\pm$0.06 &   \nodata &   \nodata    \\
73  & 3        & 105.30 &  2.08 &   6.51$\pm$0.05 &    $-$1.41$\pm$0.05 &   \nodata &   \nodata    \\
74  & 1,2,3,4  & 106.74 &  1.97 &  22.28$\pm$0.05 &    $-$2.91$\pm$0.05 &  $-$5.8$\pm$1.7 &   0.8$\pm$1.7  \\
75  & 2        & 106.40 &  1.95 &  56.00$\pm$0.05 &    $-$3.39$\pm$0.05 &   \nodata &   \nodata   \\
76  & 3        & 106.37 &  1.91 &  20.16$\pm$0.05 &     $-$23.71$\pm$0.06 &   \nodata &   \nodata   \\
77  & 1,2,3,4  & 104.43 &  1.87 &  23.93$\pm$0.05 &    $-$4.51$\pm$0.05 &  $-$9.5$\pm$1.8 &  $-$2.7$\pm$2.0   \\
78  & 3        & 106.05 &  1.87 &  $-$4.11$\pm$0.05 &    71.74$\pm$0.06 &   \nodata &   \nodata   \\
79  & 3        & 106.54 &  1.82 &   8.25$\pm$0.05 &    35.04$\pm$0.05 &   \nodata &   \nodata  \\
80  & 3        & 106.27 &  1.82 &  20.56$\pm$0.05 &     $-$65.42$\pm$0.05 &   \nodata &   \nodata   \\
81  & 3        & 105.94 &  1.81 &  $-$1.45$\pm$0.05 &    43.02$\pm$0.06 &   \nodata &   \nodata  \\
82  & 2,3      & 106.59 &  1.80 &   3.66$\pm$0.05 &     $-$45.51$\pm$0.05 &   \nodata &   \nodata   \\
83  & 1,2,3,4  &  94.90 &  1.80 & 138.95$\pm$0.05 &    67.32$\pm$0.05 &   6.7$\pm$1.2 &  14.6$\pm$1.2  \\
84  & 3        & 106.57 &  1.72 &  23.19$\pm$0.05 &     $-$46.20$\pm$0.05 &   \nodata &   \nodata   \\
85  & 4        & 105.12 &  1.67 &  38.38$\pm$0.06 &      33.81$\pm$0.06 &   \nodata &   \nodata   \\
86  & 1        & 105.73 &  1.66 &  22.67$\pm$0.05 &      $-$4.96$\pm$0.05 &   \nodata &   \nodata  \\
87  & 3        & 106.06 &  1.66 &   2.01$\pm$0.05 &      $-$1.62$\pm$0.06 &   \nodata &   \nodata   \\
88  & 3        & 106.37 &  1.64 &  20.18$\pm$0.05 &       1.15$\pm$0.05 &   \nodata &   \nodata   \\
89  & 4        & 105.42 &  1.61 &  29.71$\pm$0.06 &     $-$98.11$\pm$0.06 &   \nodata &   \nodata    \\
90  & 1,2,3,4  &  95.86 &  1.60 & 220.77$\pm$0.05 &    55.75$\pm$0.06 &  53.6$\pm$1.4 &  17.8$\pm$1.5   \\
91  & 1        & 103.89 &  1.58 &   0.44$\pm$0.05 &    $-$2.15$\pm$0.05 &   \nodata &   \nodata  \\
92  & 1,2      & 104.29 &  1.52 &  23.52$\pm$0.05 &    $-$4.52$\pm$0.05 &   \nodata &   \nodata  \\
93  & 4        & 105.12 &  1.49 &  31.87$\pm$0.06 &     $-$34.73$\pm$0.06 &   \nodata &   \nodata  \\
94  & 1,2,3    & 104.78 &  1.48 &   7.67$\pm$0.05 &      $-$1.02$\pm$0.05 & $-$13.0$\pm$3.9 & $-$12.2$\pm$4.1  \\
95  & 2,3,4    & 107.30 &  1.45 &  $-$3.86$\pm$0.05 &     1.76$\pm$0.06 & $-$27.6$\pm$2.3 &   7.0$\pm$2.4 \\
96  & 4        & 105.31 &  1.39 &  29.71$\pm$0.06 &     $-$55.31$\pm$0.06 &   \nodata &   \nodata \\
97  & 1,2,3,4  & 106.59 &  1.37 &   5.76$\pm$0.05 &      $-$2.52$\pm$0.05 & $-$22.6$\pm$1.8 &  $-$9.8$\pm$1.9   \\
98  & 3        & 103.95 &  1.35 &  $-$6.91$\pm$0.05 &   $-$38.75$\pm$0.06 &   \nodata &   \nodata   \\
99  & 2,3      & 106.78 &  1.30 &  46.15$\pm$0.05 &     0.89$\pm$0.05 &   \nodata &   \nodata   \\
100 & 4        & 104.94 &  1.28 &  30.01$\pm$0.06 &     $-$31.50$\pm$0.06 &   \nodata &   \nodata   \\
101 & 2        & 106.96 &  1.25 &  52.39$\pm$0.05 &    $-$8.39$\pm$0.05 &   \nodata &   \nodata  \\
102 & 1        & 103.88 &  1.25 &  $-$6.46$\pm$0.05 &       7.37$\pm$0.07 &   \nodata &   \nodata   \\
103 & 2,3      & 106.11 &  1.24 &  $-$0.24$\pm$0.05 &       3.29$\pm$0.06 &   \nodata &   \nodata   \\
104 & 3        & 104.07 &  1.12 &  $-$0.25$\pm$0.05 &   $-$44.67$\pm$0.06 &   \nodata &   \nodata   \\
105 & 4        & 105.11 &  1.11 &  33.36$\pm$0.06 &     2.56$\pm$0.06 &   \nodata &   \nodata   \\
106 & 4        & 105.50 &  1.11 &  37.49$\pm$0.06 &     $-$29.30$\pm$0.06 &   \nodata &   \nodata   \\
107 & 2,3      & 111.99 &  1.10 &  71.41$\pm$0.05 &     0.79$\pm$0.06 &   \nodata &   \nodata  \\
108 & 1,2,3,4  & 106.79 &  1.09 &  49.43$\pm$0.05 &      $-$3.39$\pm$0.05 &   6.3$\pm$1.7 & $-$14.1$\pm$1.7   \\
109 & 2,3      & 108.68 &  1.03 & 100.94$\pm$0.05 &      20.38$\pm$0.06 &   \nodata &   \nodata  \\
110 & 3        & 106.57 &  0.97 &   1.57$\pm$0.05 &     $-$46.93$\pm$0.05 &   \nodata &   \nodata   \\
111 & 1        & 106.80 &  0.94 &   1.44$\pm$0.05 &    $-$1.20$\pm$0.05 &   \nodata &   \nodata   \\
112 & 1,2,3    & 106.82 &  0.86 &  48.34$\pm$0.05 &    $-$3.64$\pm$0.05 &   2.5$\pm$4.0 &  $-$9.2$\pm$4.2  \\
113 & 4        & 106.91 &  0.83 &  $-$1.46$\pm$0.06 &   $-$42.03$\pm$0.06 &   \nodata &   \nodata   \\
114 & 3        & 104.67 &  0.81 &  23.33$\pm$0.05 &      $-$4.89$\pm$0.06 &   \nodata &   \nodata   \\
115 & 4        & 103.55 &  0.79 & $-$11.43$\pm$0.05 &    10.12$\pm$0.06 &   \nodata &   \nodata  \\
116 & 4        & 103.92 &  0.73 &  $-$7.58$\pm$0.06 &     8.39$\pm$0.06 &   \nodata &   \nodata  \\
117 & 1,2,3,4  & 103.66 &  0.72 & 179.75$\pm$0.05 &  $-$103.20$\pm$0.05 &  18.9$\pm$2.2 &   1.3$\pm$2.4   \\
118 & 1,2      &  94.32 &  0.69 & 139.93$\pm$0.05 &      66.86$\pm$0.05 &   \nodata &   \nodata   \\
119 & 4        & 103.64 &  0.68 &   0.34$\pm$0.06 &  $-$114.27$\pm$0.06 &   \nodata &   \nodata   \\
120 & 3,4      & 107.47 &  0.66 & 167.34$\pm$0.05 &  $-$261.03$\pm$0.06 &   \nodata &   \nodata   \\
121 & 1,2,3    & 107.35 &  0.61 &  50.20$\pm$0.05 &      $-$3.27$\pm$0.05 &  $-$0.6$\pm$3.7 &  $-$9.6$\pm$4.0  \\
122 & 1        & 107.18 &  0.57 &  $-$3.75$\pm$0.05 &       0.95$\pm$0.05 &   \nodata &   \nodata  \\
123 & 1,2,3    &  94.48 &  0.57 & 214.01$\pm$0.05 &    54.68$\pm$0.06 &  48.1$\pm$2.6 &  18.7$\pm$2.8 \\
124 & 3,4      & 107.11 &  0.53 &  $-$6.35$\pm$0.05 &    6.41$\pm$0.07  &   \nodata &   \nodata  \\
125 & 2        &  95.24 &  0.49 & 202.75$\pm$0.05 &     51.60$\pm$0.05  &   \nodata &   \nodata  \\
126 & 1        & 107.01 &  0.48 &  $-$5.96$\pm$0.05 &    6.03$\pm$0.05  &   \nodata &   \nodata  \\
127 & 1,2,3,4  &  94.75 &  0.41 & 206.85$\pm$0.05 &     51.60$\pm$0.06  &  43.1$\pm$1.5 &  14.0$\pm$1.6   \\
128 & 4        & 108.82 &  0.38 &  $-$1.59$\pm$0.06 &    3.18$\pm$0.06  &   \nodata &   \nodata   \\
129 & 1,2,3,4  &  93.64 &  0.38 & 140.63$\pm$0.05 &     66.57$\pm$0.05  &   5.9$\pm$1.6 &  14.3$\pm$1.6   \\
130 & 3        & 103.41 &  0.38 &  $-$6.86$\pm$0.06 &   74.45$\pm$0.07  &   \nodata &   \nodata   \\
131 & 2,3,4    & 107.11 &  0.37 &  84.03$\pm$0.05 &     $-$0.53$\pm$0.06  &  $-$0.8$\pm$2.7 & $-$15.5$\pm$2.9  \\
132 & 2        &  95.94 &  0.36 & 138.45$\pm$0.05 &     67.75$\pm$0.06  &   \nodata &   \nodata   \\
133 & 1,2,3    &  93.60 &  0.36 & 213.24$\pm$0.05 &     55.03$\pm$0.11  &  46.1$\pm$2.9 &  21.3$\pm$3.1  \\
134 & 1,2      &  99.32 &  0.34 & 121.98$\pm$0.05 &     91.75$\pm$0.06  &   \nodata &   \nodata   \\
135 & 1,2,3,4  &  93.62 &  0.34 & 205.00$\pm$0.05 &     51.40$\pm$0.06  &  42.0$\pm$1.7 &  11.4$\pm$1.7   \\
136 & 1,2,3,4  &  95.47 &  0.34 & 204.63$\pm$0.06 &     51.38$\pm$0.06  &  52.1$\pm$1.8 &  17.2$\pm$2.1   \\
137 & 1,2,3    & 109.59 &  0.33 &  98.29$\pm$0.05 &     11.15$\pm$0.06  & $-$12.3$\pm$3.0 & $-$12.6$\pm$3.4  \\
138 & 3,4      &  99.16 &  0.31 & 122.29$\pm$0.05 &     91.99$\pm$0.06  &   \nodata &   \nodata  \\
139 & 4        &  93.97 &  0.31 & 214.80$\pm$0.06 &     55.26$\pm$0.06  &   \nodata &   \nodata  \\
140 & 3        & 103.39 &  0.30 &  58.58$\pm$0.06 &     $-$3.38$\pm$0.07  &   \nodata &   \nodata   \\
141 & 3        & 107.37 &  0.28 &  65.53$\pm$0.05 &    5.02$\pm$0.06  &   \nodata &   \nodata \\
142 & 1,2      &  95.14 &  0.27 & 202.97$\pm$0.05 &     49.48$\pm$0.05  &   \nodata &   \nodata   \\
143 & 1,2,3    & 107.23 &  0.25 & 166.00$\pm$0.05 & $-$257.48$\pm$0.06  &   \nodata &   \nodata   \\
144 & 1,2,3    & 108.39 &  0.25 &  45.92$\pm$0.05 &     $-$6.11$\pm$0.05  &   2.7$\pm$3.7 &  $-$8.6$\pm$4.4  \\
145 & 4        &  95.90 &  0.25 & 218.01$\pm$0.06 &     61.98$\pm$0.07  &   \nodata &   \nodata  \\
146 & 1,2,3    &  94.16 &  0.23 & 136.39$\pm$0.05 &     74.76$\pm$0.06  &   6.9$\pm$1.9 &   7.9$\pm$2.2   \\
147 & 1,2,3    & 111.87 &  0.22 & 171.96$\pm$0.05 &  $-$42.98$\pm$0.05  &  20.3$\pm$2.3 &  $-$6.5$\pm$2.6  \\
148 & 1,2      &  99.12 &  0.22 & 122.43$\pm$0.05 &     91.64$\pm$0.05  &   \nodata &   \nodata  \\
149 & 4        & 107.41 &  0.21 &  $-$8.25$\pm$0.06 &    7.57$\pm$0.07  &   \nodata &   \nodata \\
150 & 4        & 103.23 &  0.21 &  $-$6.00$\pm$0.07 &  $-$22.28$\pm$0.07  &   \nodata &   \nodata  \\
151 & 2        &  95.20 &  0.19 & 135.49$\pm$0.05 &     77.14$\pm$0.07  &   \nodata &   \nodata \\
152 & 4        & 108.66 &  0.19 &   2.04$\pm$0.07 &     $-$4.78$\pm$0.07  &   \nodata &   \nodata  \\
153 & 1,2      & 102.57 &  0.16 & 100.89$\pm$0.05 &    8.93$\pm$0.05  &   \nodata &   \nodata \\
154 & 1,2,3    &  95.22 &  0.16 & 202.79$\pm$0.06 &     56.68$\pm$0.05  &  46.4$\pm$3.0 &   1.8$\pm$3.7 \\
155 & 2,3,4    &  93.33 &  0.16 & 205.03$\pm$0.06 &     58.09$\pm$0.06  &  27.8$\pm$4.5 &  $-$1.8$\pm$4.5  \\
156 & 1        &  99.90 &  0.14 &  99.55$\pm$0.05 &      9.10$\pm$0.06  &   \nodata &   \nodata  \\
157 & 4        & 103.02 &  0.14 &  $-$2.92$\pm$0.07 &   $-$0.52$\pm$0.08  &   \nodata &   \nodata  \\
158 & 1,2,3    &  96.62 &  0.13 & 217.56$\pm$0.05 &     54.15$\pm$0.06  &  51.8$\pm$2.0 &  18.0$\pm$2.4 \\
159 & 1        &  95.29 &  0.13 & 202.24$\pm$0.06 &     47.80$\pm$0.06  &   \nodata &   \nodata  \\
160 & 2        & 107.64 &  0.13 & 182.83$\pm$0.05 & $-$174.07$\pm$0.07  &   \nodata &   \nodata  \\
161 & 3        & 108.71 &  0.13 &   2.47$\pm$0.06 &     $-$4.01$\pm$0.05  &   \nodata &   \nodata   \\
162 & 2        & 107.42 &  0.12 & 170.05$\pm$0.06 & $-$263.08$\pm$0.08  &   \nodata &   \nodata   \\
163 & 4        &  94.87 &  0.12 & 135.39$\pm$0.06 &     73.40$\pm$0.08  &   \nodata &   \nodata   \\
164 & 2        &  94.98 &  0.12 & 201.51$\pm$0.06 &     43.74$\pm$0.08  &   \nodata &   \nodata   \\
165 & 3        & 108.47 &  0.12 &  82.90$\pm$0.06 &      1.27$\pm$0.09  &   \nodata &   \nodata   \\
166 & 3        &  95.84 &  0.11 & 225.25$\pm$0.06 &      7.28$\pm$0.10  &   \nodata &   \nodata   \\
167 & 2        &  98.34 &  0.11 & 128.73$\pm$0.06 &     86.69$\pm$0.09  &   \nodata &   \nodata   \\
168 & 3        & 107.50 &  0.10 &  $-$1.99$\pm$0.07 &    2.06$\pm$0.06  &   \nodata &   \nodata   \\
\enddata
\tablecomments{Table~\ref{tbl:water} in printer. Column~1 gives the
feature label number; column~2 lists the observing epochs at which
the feature was detected; columns~3~and~4 report the value of the
intensity-weighted LSR velocity and flux density of the strongest
spot, averaged over the observing epochs; columns~5~and~6 give the
position offsets (with the associated errors) along the R.A. and
Dec. axes relative to the feature \#1, measured at the first epoch
of detection; columns~7~and~8 give the components of the relative
proper motion (with the associated errors) along the R.A. and Dec.
axes, measured
with respect to the reference feature \#0.\\
}
\end{deluxetable}

\end{document}